\documentclass[10pt,a4paper,twoside]{article}
\usepackage{epsfig}
\usepackage{baltlat6}
\usepackage{array}
\usepackage{here}
\begin{document}
\ \
\vspace{0.5mm}
\setcounter{page}{1}
\vspace{8mm}

\titlehead{Baltic Astronomy, in press}

\titleb{The X-ray Evolution of the Symbiotic Star V407 Cygni during
  its 2010 Outburst}

\begin{authorl}
\authorb{K. Mukai}{1,2}
\authorb{T. Nelson}{1,2,3}
\authorb{L. Chomiuk}{4,5,6}
\authorb{D. Donato}{7,8} and
\authorb{J. Sokoloski}{9}
\end{authorl}

\begin{addressl}
\addressb{1}{CRESST and X-ray Astrophysics Laboratory, NASA/GSFC,
  Greenbelt, MD 20771, USA}
\addressb{2}{Department of Physics, University of Maryland, Baltimore
  County, 1000 Hilltop Circle, Baltimore, MD 21250, USA}
\addressb{3}{Present Address: School of Physics and Astronomy,
  University of Minnesota, 115 Church St SE, Minneapolis, MN 55455,
  USA}
\addressb{4}{Harvard-Smithsonian Center for Astrophysics, 60 Garden
  St., Cambridge, MA 02138, USA}
\addressb{5}{National Radio Astronomy Observatory, P.O. Box 0,
  Socorro, MN 87801, USA}
\addressb{6}{Jansky Fellow}
\addressb{7}{CRESST and Astroparticle Physics Laboratory, NASA/GSFC,
  Greenbelt, MD 20771, USA}
\addressb{8}{Department of Astronomy, University of Maryland,
 College Park, MD 20742, USA}
\addressb{9}{Columbia Astrophysics Laboratory, 550 W. 120th St., 1027
  Pupin Hall, Columbia University, New York, NY 10027, USA}
\end{addressl}

\submitb{Received: 2011 September 1}

\begin{summary} 
We present a summary of {\sl Swift\/} and {\sl Suzaku\/} X-ray
observations of the 2010 nova outburst of the symbiotic star, V407
Cyg.  The {\sl Suzaku\/} spectrum obtained on day 30 indicates the
presence of the supersoft component from the white dwarf surface, as
well as optically thin component from the shock between the nova
ejecta and the Mira wind.  The {\sl Swift\/} observations then allow
us to track the evolution of both components from day 4 to day 150.
Most notable is the sudden brightening of the optically thin
component around day 20.  We identify this as the time when the blast
wave reached the immediate vicinity of the photosphere of the Mira.
We have developed a simple model of the blast wave--wind
interaction that can reproduce the gross features of the X-ray
evolution of V407 Cyg, and explore a parameter space of ejected mass, binary separation and Mira mass loss rate..  If the model is correct, the binary separation
is likely to be larger then previously suggested and the mass loss rate of the Mira is likely to
be relatively low.
\end{summary}

\begin{keywords} stars: white dwarfs --X-rays: stars\end{keywords}

\resthead{X-ray Evolution of V407 Cyg during its 2010 Outburst}
{K. Mukai et al.}

\sectionb{1}{INTRODUCTION}

The 2010 outburst of the little-studied symbiotic star, V407 Cygni,
was detected on 2010 March 10.  Subsequent optical photometry and
spectroscopy (Munari et al. 2011; Shore et al. 2011) clearly establish
this as a classical nova event with t$_2$=5.9 d and t$_3$=24 d, unlike
the smaller amplitude events seen in 1936 and 1998.  Thus, the closest
analogue to the 2010 outburst of V407 Cyg  is the 2006 outburst of the
recurrent nova RS Ophiuchi.  In the latter case, strong hard X-ray
emission was almost immediately discovered and intensively monitored
using {\sl RXTE\/} and {\sl Swift\/} (Sokoloski et al. 2006; Bode et
al. 2006).  In the subsequent months, RS Oph became a bright supersoft
X-ray source; its evolution was followed with {\sl Swift\/} and high
resolution, high signal-to-noise spectra were obtained using {\sl
XMM-Newton\/} and {\sl Chandra\/} grating instruments (e.g., Nelson et
al. 2008; Ness et al. 2009; Osborne et al. 2011).

The X-ray emission of V407 Cyg evolved rather differently.  Its hard
X-ray emission developed more slowly and peaked at a much lower
level than in RS Oph, and there was no obvious and bright supersoft
phase, as was shown in the analysis of the {\sl Swift\/} data (Shore
et al. 2011).   The likely reason for this is the different binary
parameters.  RS Oph is an S-type symbiotic with a normal red giant
mass donor with an orbital period of 455.72 d (Zajczyk et al. 2007)
and an orbital separation of $\sim$1 AU; a 3,000 km\,s$^{-1}$ nova
blast wave would reach the vicinity of the red giant in half a day,
get strongly shocked, and become a luminous hard X-ray source
(see, e.g., Orlando et al. 2009).  V407 Cyg, on the other hand,
is a D-type symbiotic with a Mira type mass donor.  Munari et
al. (1990) noted dust obscuration events separated by 43 years and
proposed that as the orbital period.  While plausible, we do not
consider this to be firmly established.  Nevertheless, the current
body of data on V407 Cyg indicates a substantially wider binary
separation than for RS Oph, of at least 10 AU. Therefore, the nova
blast wave propagates in a lower density environment, and it takes
significantly longer to reach the Mira than was the case for RS Oph.

Most remarkably, V407 Cyg was detected as a transient GeV
$\gamma$-ray source with {\sl Fermi\/} LAT (Abdo et al. 2010).
The first significant detection was obtained on 2010 March 10, the
day of the optical discovery, and the last significant detection was
on day 15 of the outburst (2010 March 25).  As was already pointed
out by Abdo et al. (2010), the fading of the $\gamma$-rays coincided
with the rapid brightening of the X-rays.  This provides additional
motivation for our effort to understand the X-ray evolution of V407 Cyg.

\sectionb{2}{SUMMARY OF X-RAY OBSERVATIONS AND RESULTS}


\begin{figure}[!tH]
\vbox{
\centerline{\psfig{figure=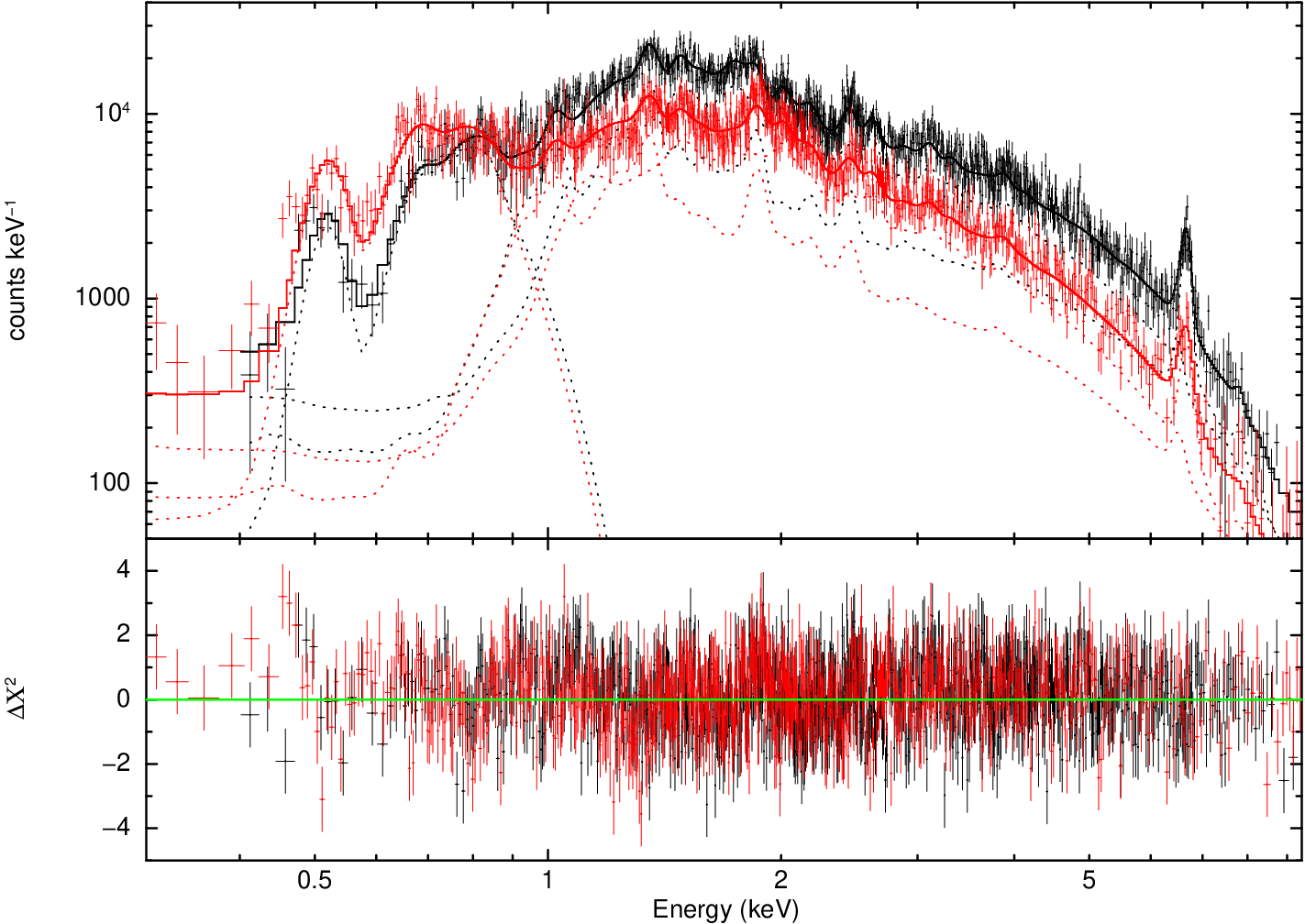,width=75mm,angle=270,clip=}}
\vspace{1mm}
\captionb{1}
{The {\sl Suzaku\/} X-ray spectra (black: XIS0+XIS3; red: XIS1) of
  V407 Cyg obtained on day 30, fitted with a two component (one in
  collisional ionization equilibrium and the other not) plasma model
  plus a supersoft blackbody.}
}
\end{figure}


X-ray observations of V407 Cyg were obtained with {\sl Swift\/} XRT
between day 3.8 and day 142 (34 visits, typical exposures of 1--5 ks).
We also obtained a 42 ks {\sl Suzaku\/} target-of-opportunity
observation on day 30, near the peak of the X-ray emission.  We have
performed a detailed spectral analysis of the {\sl Suzaku\/} data, and
re-analyzed the {\sl Swift\/} data in light of the {\sl Suzaku\/}
results.  Full details will be published elsewhere (Nelson et
al. 2012).   Here, we
provide a summary of key findings, before exploring a range of
parameters for the Mira, the binary, and the nova ejecta in the next section.

The {\sl Suzaku\/} spectra (one for the two units of XIS with front
illuminated CCDs combined, and one for the XIS unit with a back
illuminated chip)  on Day 30 (Figure 1) consist of a soft
component, dominant below $\sim$0.9 keV, and a hard component. 
The hard component is optically thin with kT of order 3 keV, and
exhibits prominent lines of He-like Fe, S, Si, and Mg.  While the
He-like Fe line is expected in a
$\sim$3 keV thermal plasma in collisional ionization equilibrium
(CIE),  the other lines are not.  They are expected of 3 keV plasmas
on their way toward ionization equilibrium (Smith \& Hughes 2010).
Nelson et al. (2012) were able to obtain a satisfactory fit above 1 keV
by using a mixture of CIE and non ionization equilibrium (NIE)
plasmas, and interpreted this as being due to the shock near the Mira
(higher density environment) and the shock away from the Mira
(lower density, and hence did not have the time to reach ionization
equilibrium). 


\begin{figure}[!tH]
\vbox{
\centerline{\psfig{figure=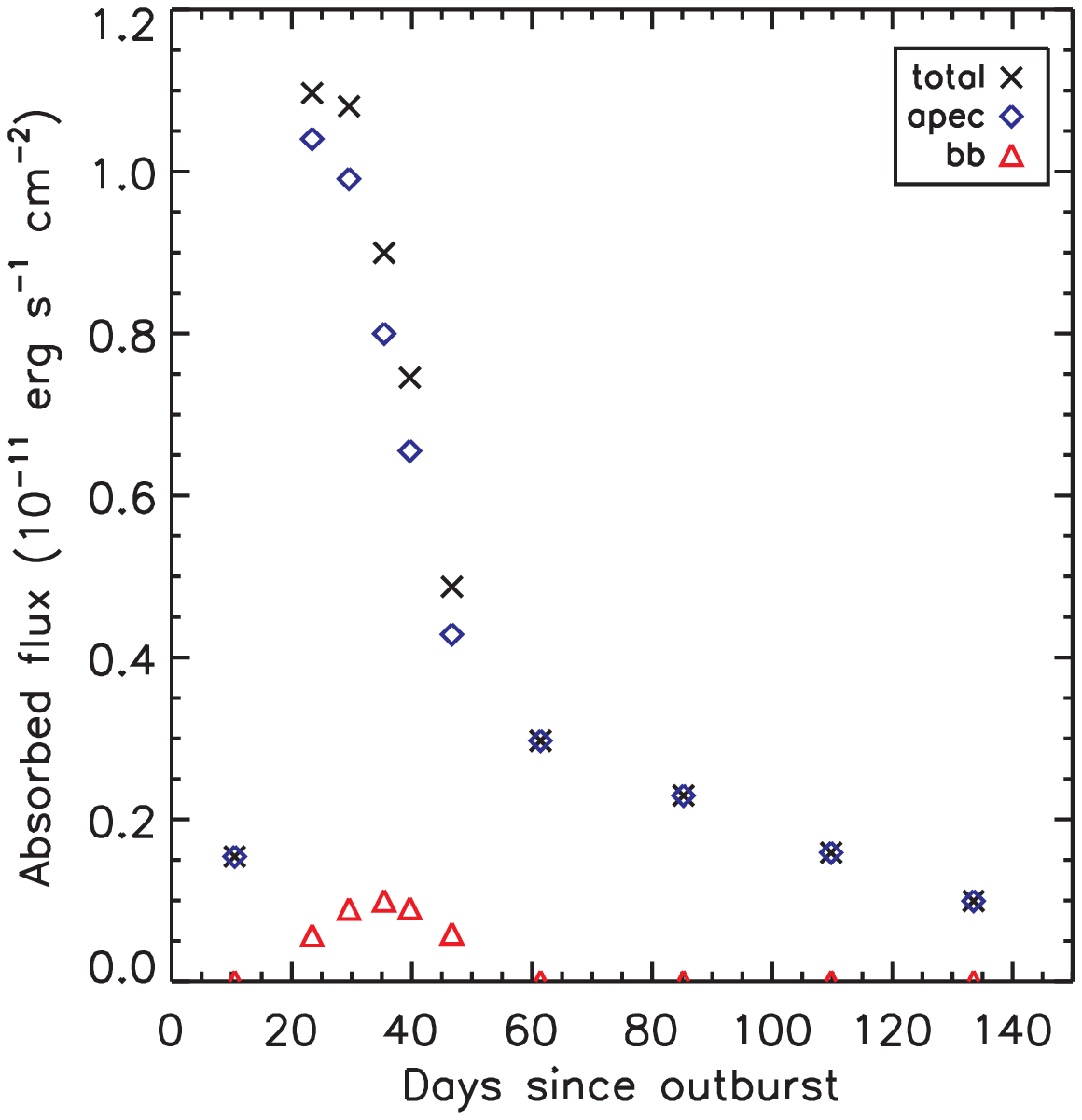,width=80mm,angle=0,clip=}}
\vspace{1mm}
\captionb{2}
{The X-ray flux evolution of the hard (APEC) and soft (blackbody)
  components of V407 Cyg, as inferred from {\sl Swift\/} XRT data.}
}
\end{figure}

 
The shape of the soft component is poorly constrained by the data.
However, fits using optically thin plasma models require an
unrealistically high emission measure; therefore, we discard this
possibility as implausible.  We do achieve a reasonable fit using a
blackbody to represent the soft component, with a best-fit luminosity
of 8.4$\times$10$^{39}$ erg\,s$^{-1}$.  We interpret this as the
supersoft emission from the surface of the nuclear burning white
dwarf, whose luminosity is overestimated due to the use of the
blackbody model.

Individual {\sl Swift\/} XRT spectra have far fewer counts than the
{\sl Suzaku\/} data, even though we analyzed sliding sums of
3 neighboring pointings.  We therefore used a simple spectral
model, consisting of an absorbed blackbody and a single APEC optically
thin, CIE, plasma model.  By applying the same model consistently to
all the {\sl Swift\/} data, we were able to infer the evolution of
spectral parameters throughout the outburst.  We show the
X-ray flux of the soft and hard components in Figure 2. 

The blackbody component is detected only during days 20--50.
The rapid evolution of the supersoft emission and the relatively high
temperature (kT=40--60 eV) both suggest that V407 Cyg contains a massive white
dwarf.  It was not seen to make a dominant contribution to the
observed X-ray flux during any period because of high intrinsic
absorption in the Mira wind.  

The faint, hard emission observed during the
first two weeks is consistent with a $\sim$6 keV plasma with high
($>$10$^{23}$ cm$^{-2}$) absorption.  Between day 20 and 50, the
emission measure, kT, and N$_H$ all declined sharply, and thereafter
they declined at a much reduced rate.

\sectionb{3}{EXPLORATION OF PARAMETER SPACE}


\begin{figure}[!tH]
\vbox{
\centerline{
\begin{tabular}{ccc}
\hspace{1mm}
\psfig{figure=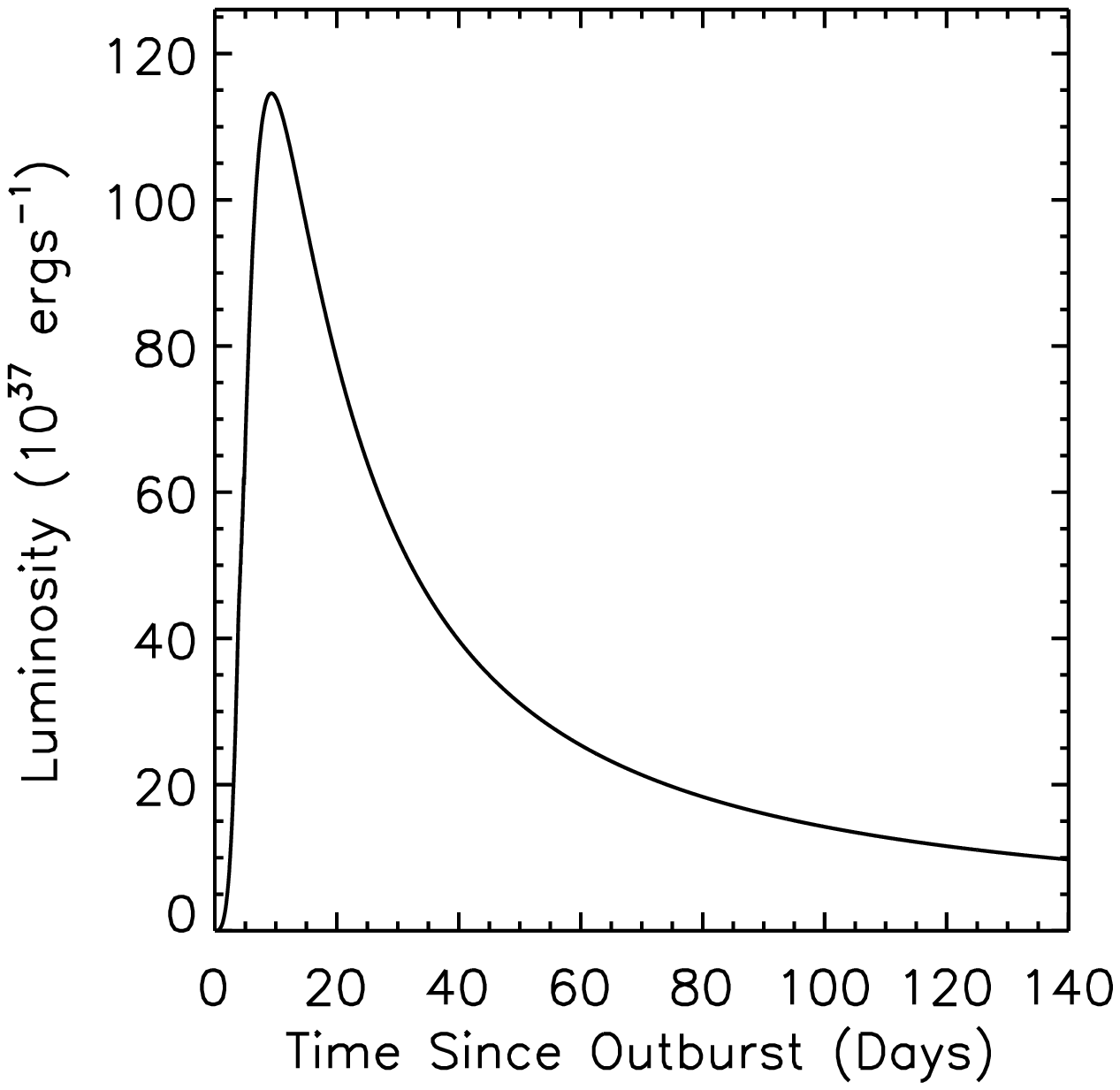,width=39mm,angle=90} &
\hspace{-2mm}
\psfig{figure=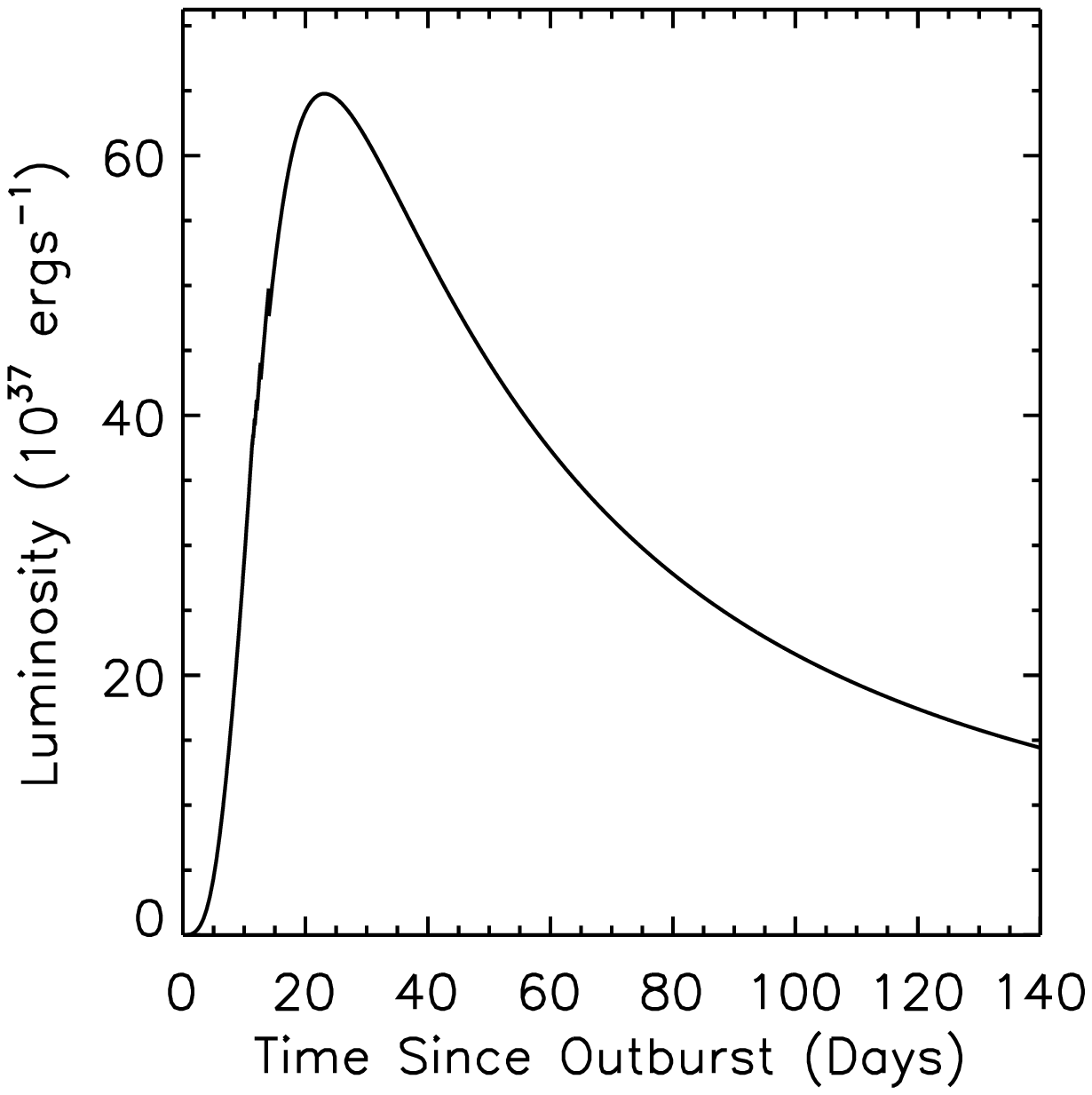,width=39mm,angle=90} &
\hspace{-2mm}
\psfig{figure=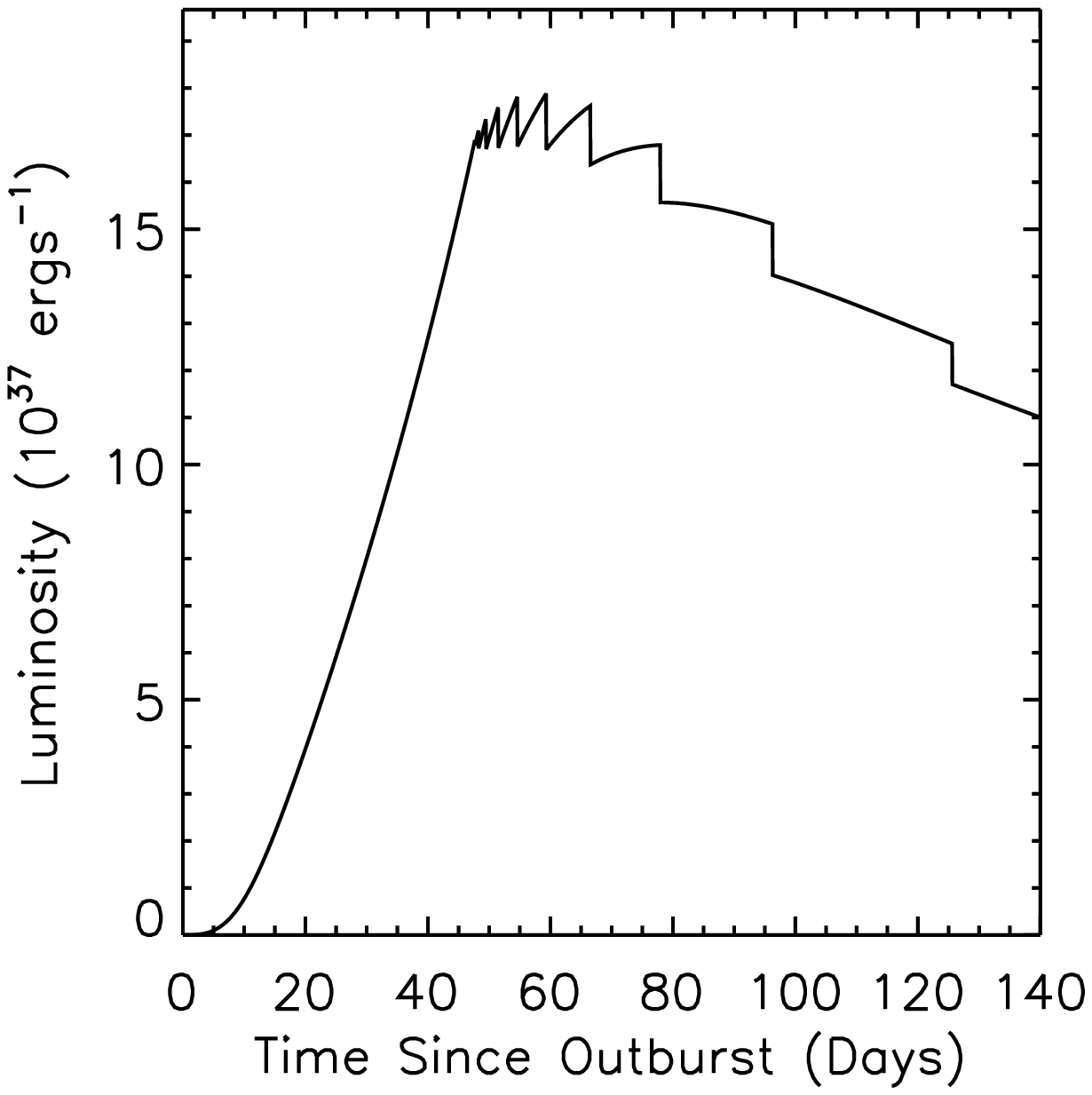,width=39mm,angle=90} \\
\hspace{1mm}
\psfig{figure=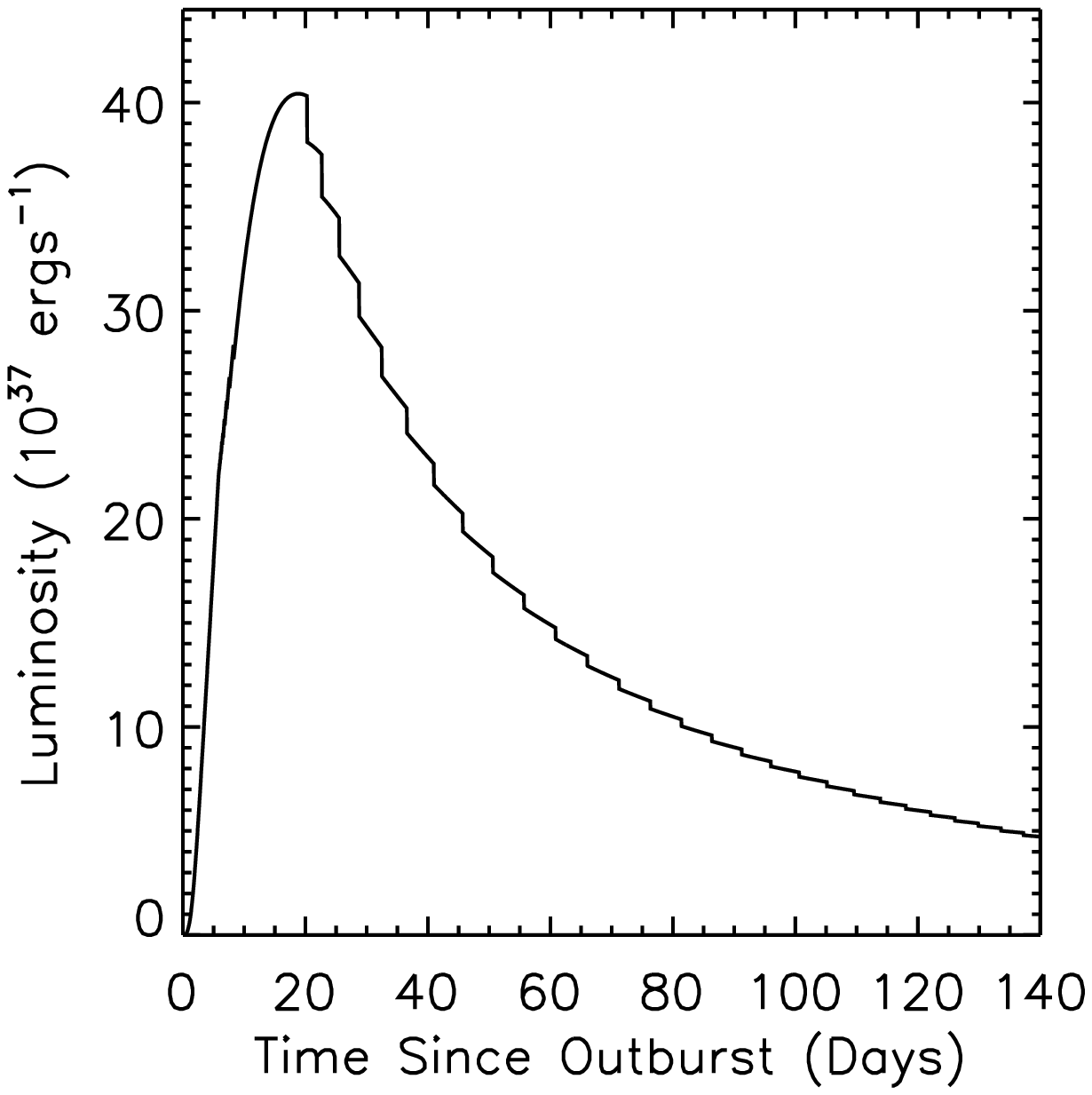,width=39mm,angle=90} &
\hspace{-2mm}
\psfig{figure=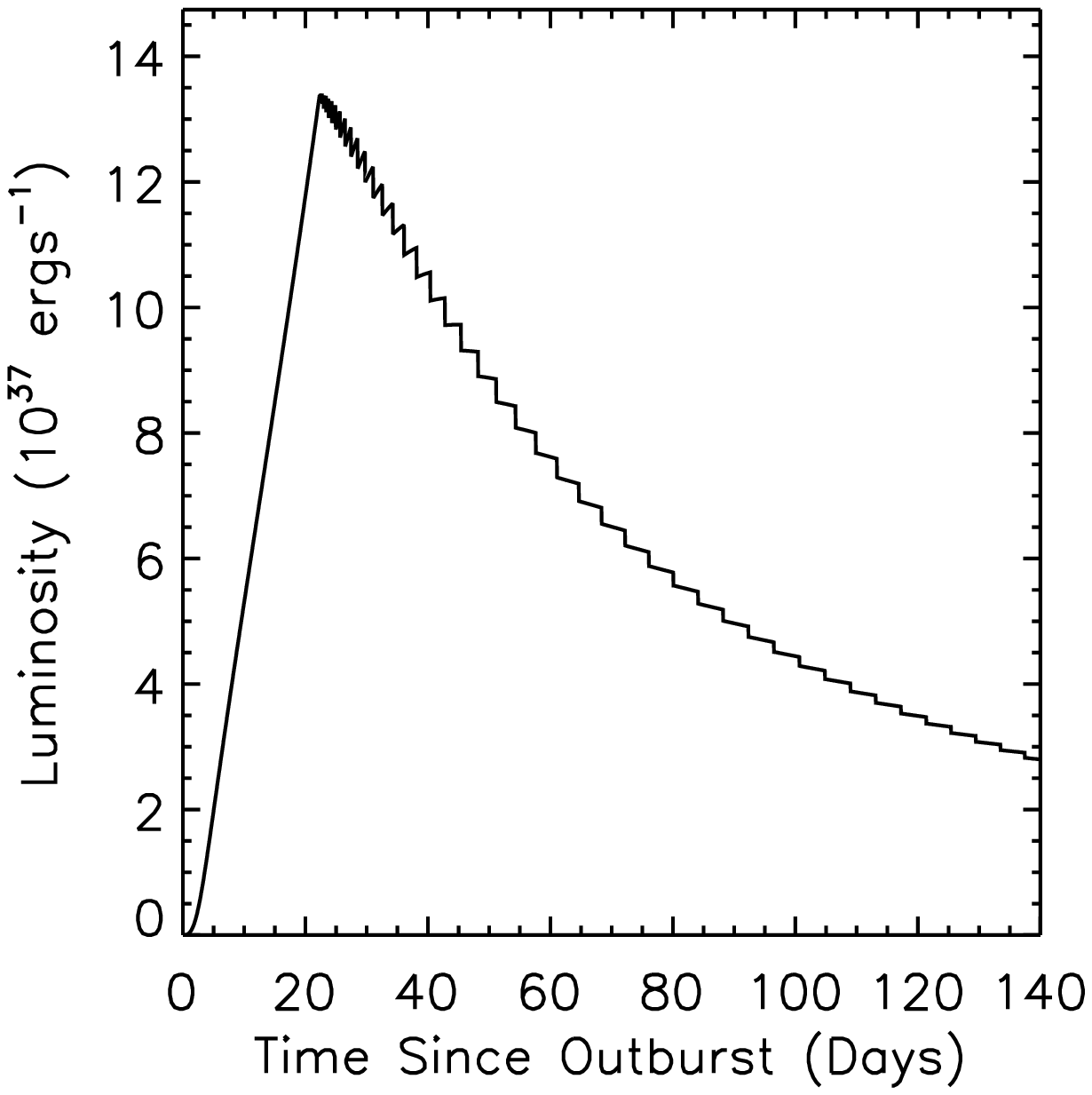,width=39mm,angle=90}
&
\hspace{-2mm}
\psfig{figure=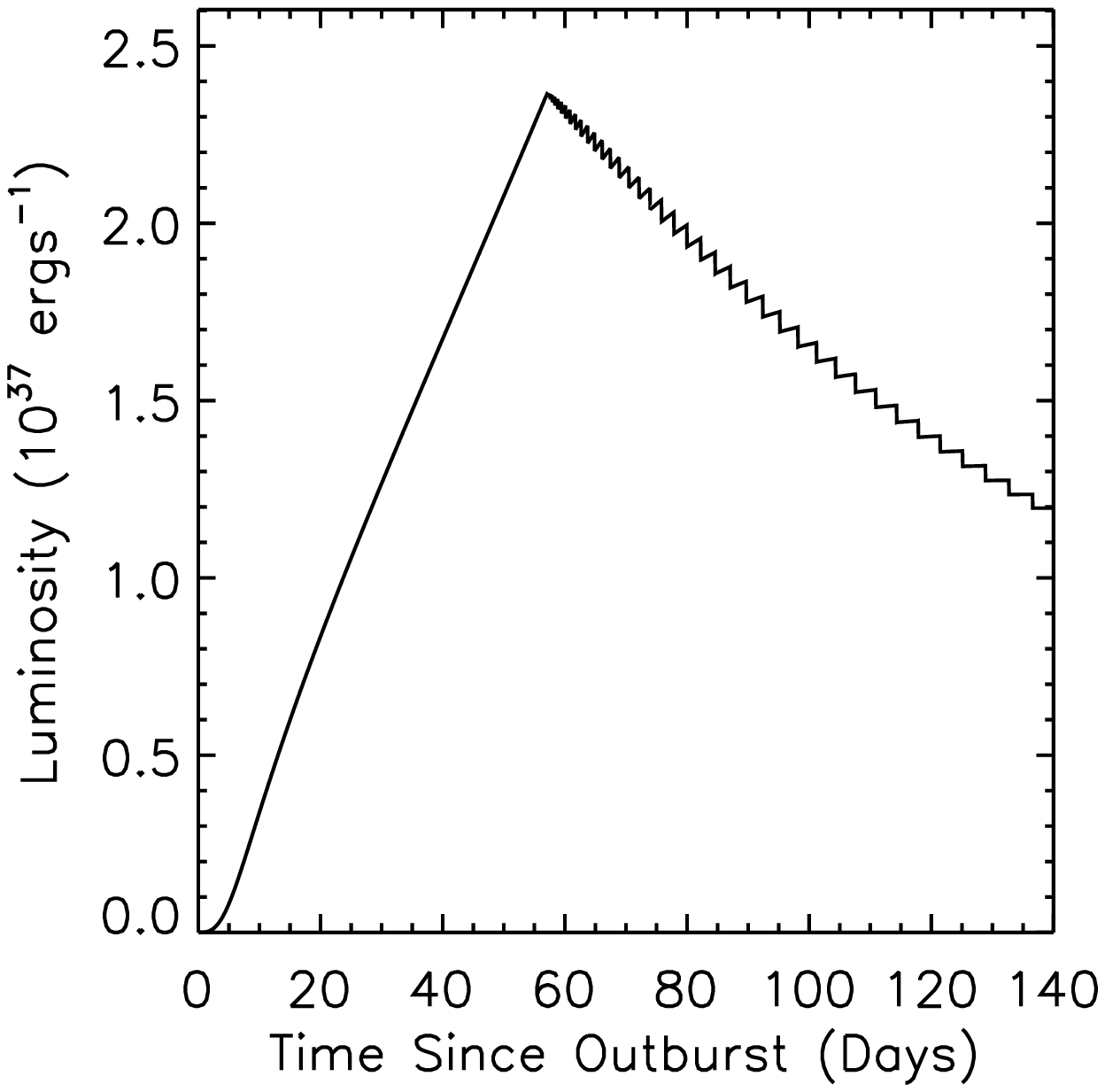,width=39mm,angle=90} \\
\hspace{1mm}
\psfig{figure=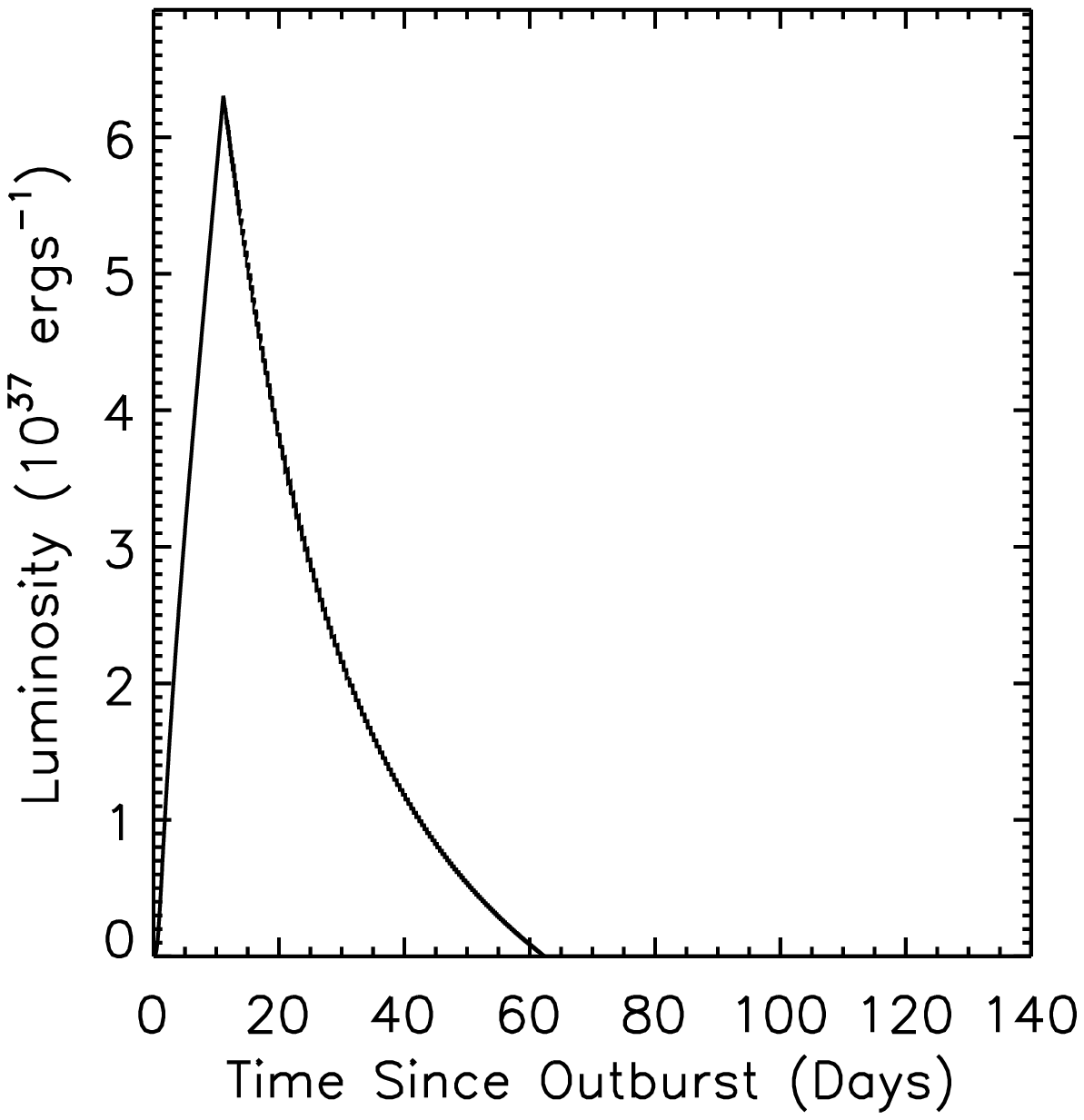,width=39mm,angle=90} &
\hspace{-2mm}
\psfig{figure=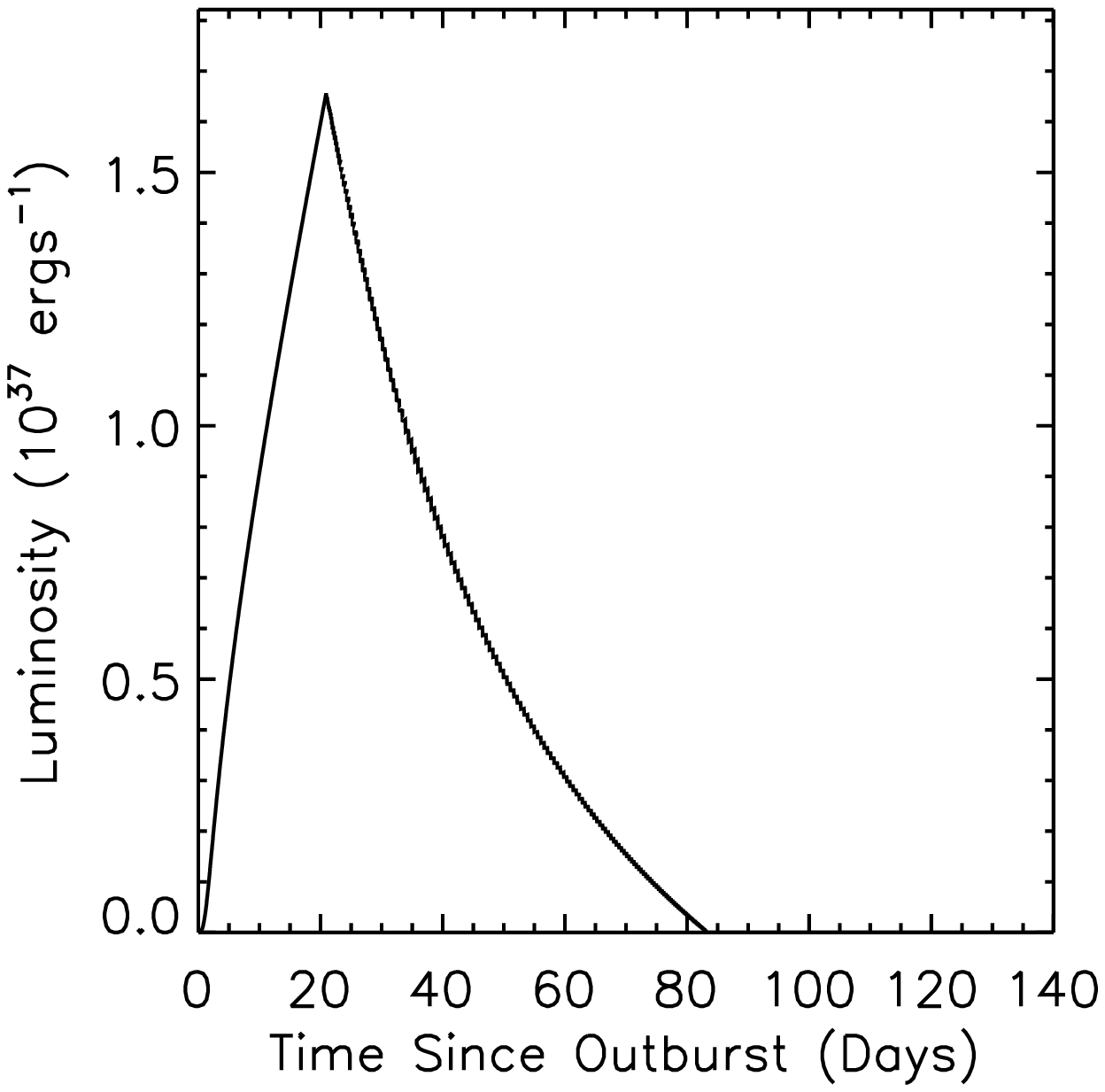,width=39mm,angle=90} &
\hspace{-2mm}
\psfig{figure=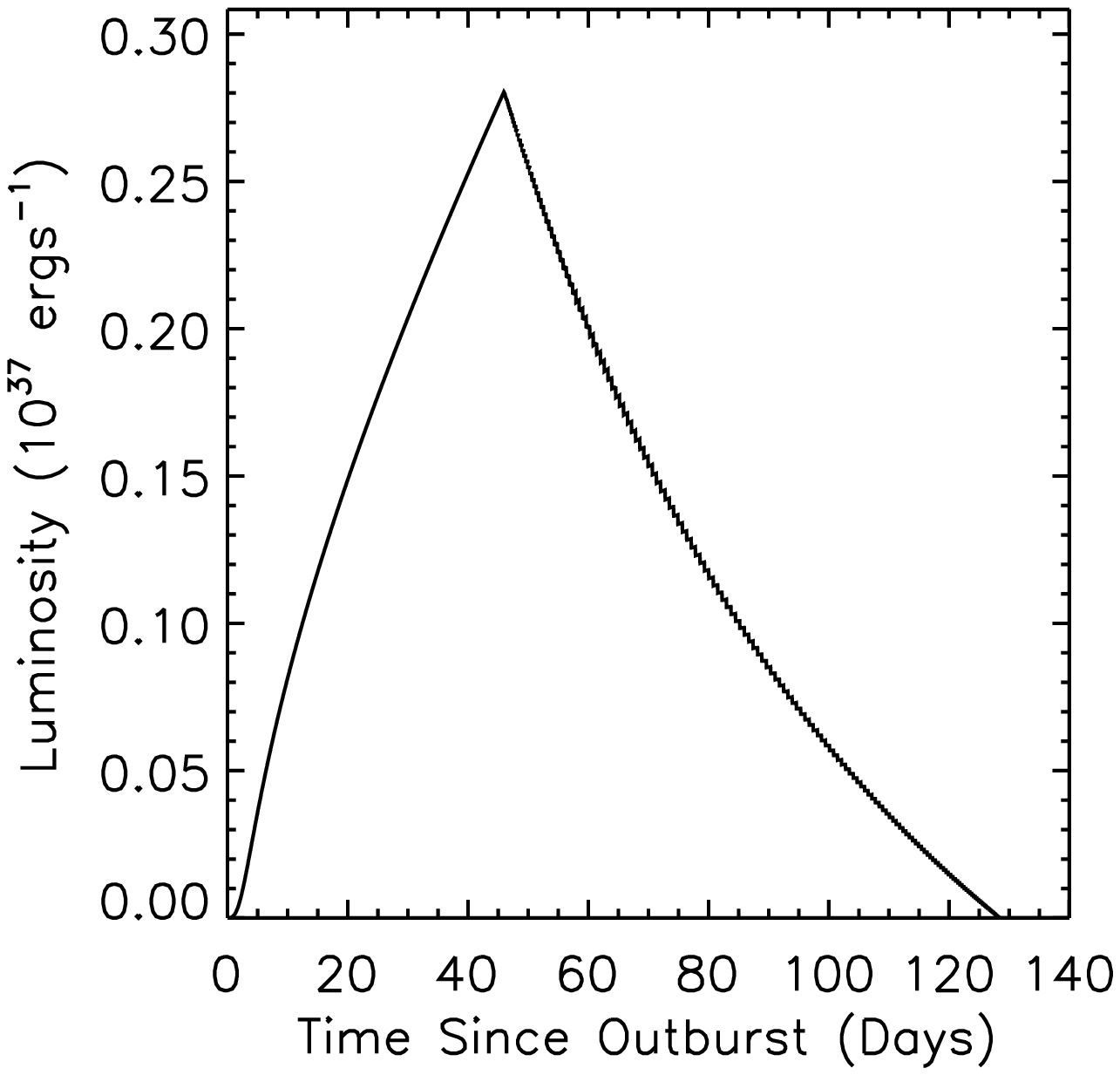,width=39mm,angle=90}
\end{tabular}}
\vspace{1mm}
\captionb{3}
{Luminosity evolution over the first 140 days predicted by our simple model
  for an assumed Mira mass loss rate of 10$^{-5}$
  M$_\odot$\,yr$^{-1}$.  The top, middle, and bottom rows are the
  results for nova ejecta masses of 10$^{-5}$, 10$^{-6}$,  and
  10$^{-7}$ M$_\odot$, respectively, and the left, middle, and right
  columns are for binary separation of 10, 20, and 50 R$_\odot$. The stepwise features in some of the plots 
  are an artifact of our simple model - if the post-shock temperature of a given region is less than 10$^6$ K, we assume it no 
  longer emits X-rays.  Therefore, discontinuities in the luminosity arise as material is strongly decelerated.  }
}
\end{figure}


The most likely origin of the hard component of the X-ray spectrum of
V407 Cyg is the forward shock driven into the Mira wind by the nova
blast wave.  We have therefore developed a simple model of this
interaction (full details are presented in Nelson et al., 2012).  The model tracks the sweeping up of the Mira wind by
the ejecta as they travel outwards from the white dwarf, and estimates
the global properties of the resulting X-ray emission.   We assume that
the Mira wind follows an $r^{-2}$ density profile, and adopt a
photospheric radius of 500 R$_\odot$ and a wind velocity of 30
km\,s$^{-1}$.  The wind mass loss rate is a key parameter and is
poorly constrained in V407 Cyg: we therefore compare the results for
three assumed values, 10$^{-7}$, 10$^{-6}$,  and
10$^{-5}$ M$_\odot$\,yr$^{-1}$.

The white dwarf is significantly offset from the center of the Mira
wind, by the orbital separation $a$ that is also unknown.  We explore
the effects of different binary separations by running models for $a$=10,
20, and 50 R$_\odot$.  Based on the full width at zero intensity of
the Balmer lines (Munari et al. 2011; Shore et al 2011), we adopt the initial
ejecta velocity of 3200 km\,s$^{-1}$.  If the blast wave is undecelerated,
it would take 4, 10, and 25 days to reach the surface of the Mira for
these three values of $a$, respectively, after subtracting the red
giant radius of 500 R$_\odot$ ($\sim$2.3 AU).


\begin{figure}[!tH]
\vbox{
\centerline{
\begin{tabular}{ccc}
\hspace{1mm}
\psfig{figure=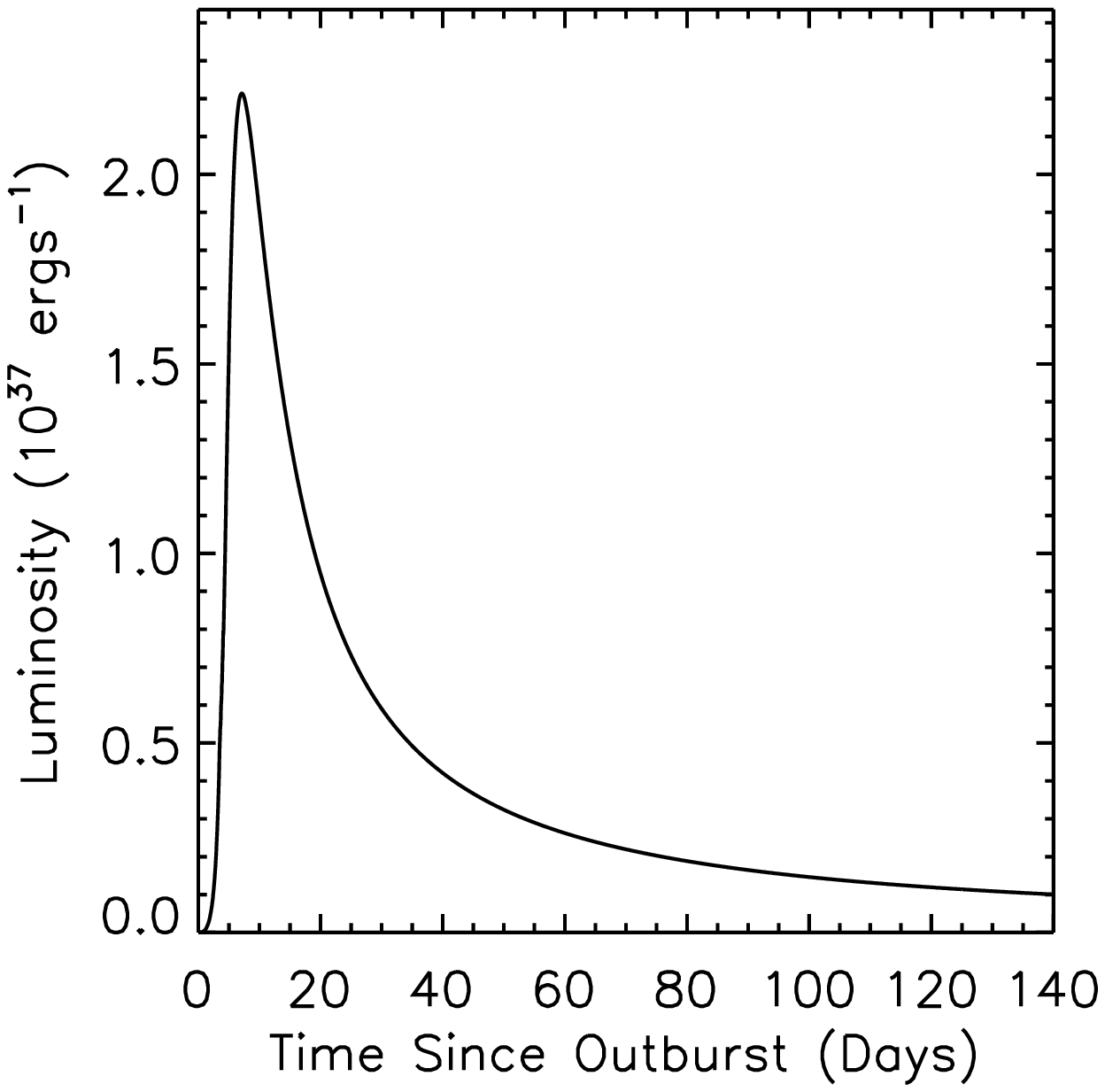,width=39mm,angle=90} &
\hspace{-2mm}
\psfig{figure=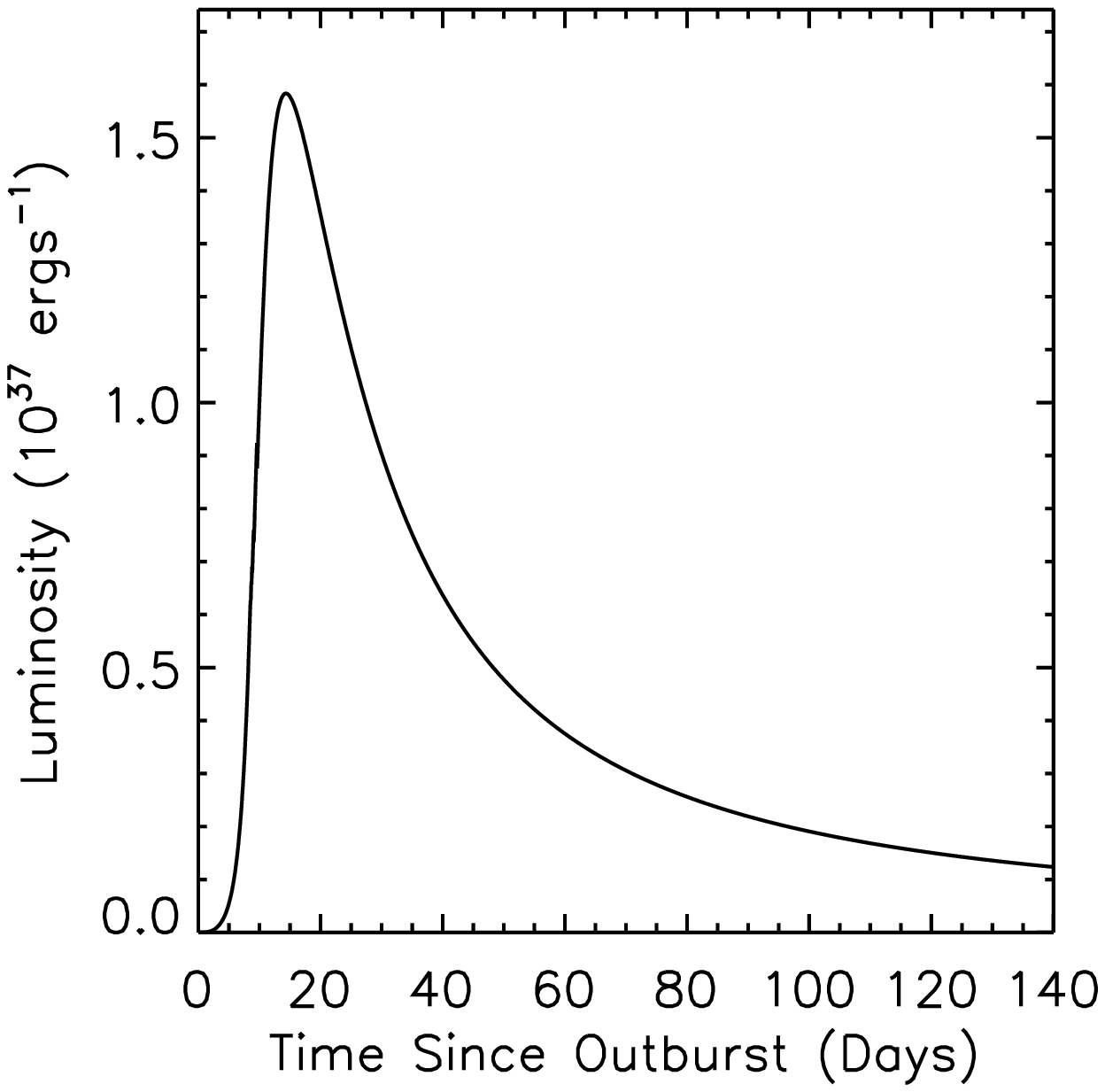,width=39mm,angle=90} &
\hspace{-2mm}
\psfig{figure=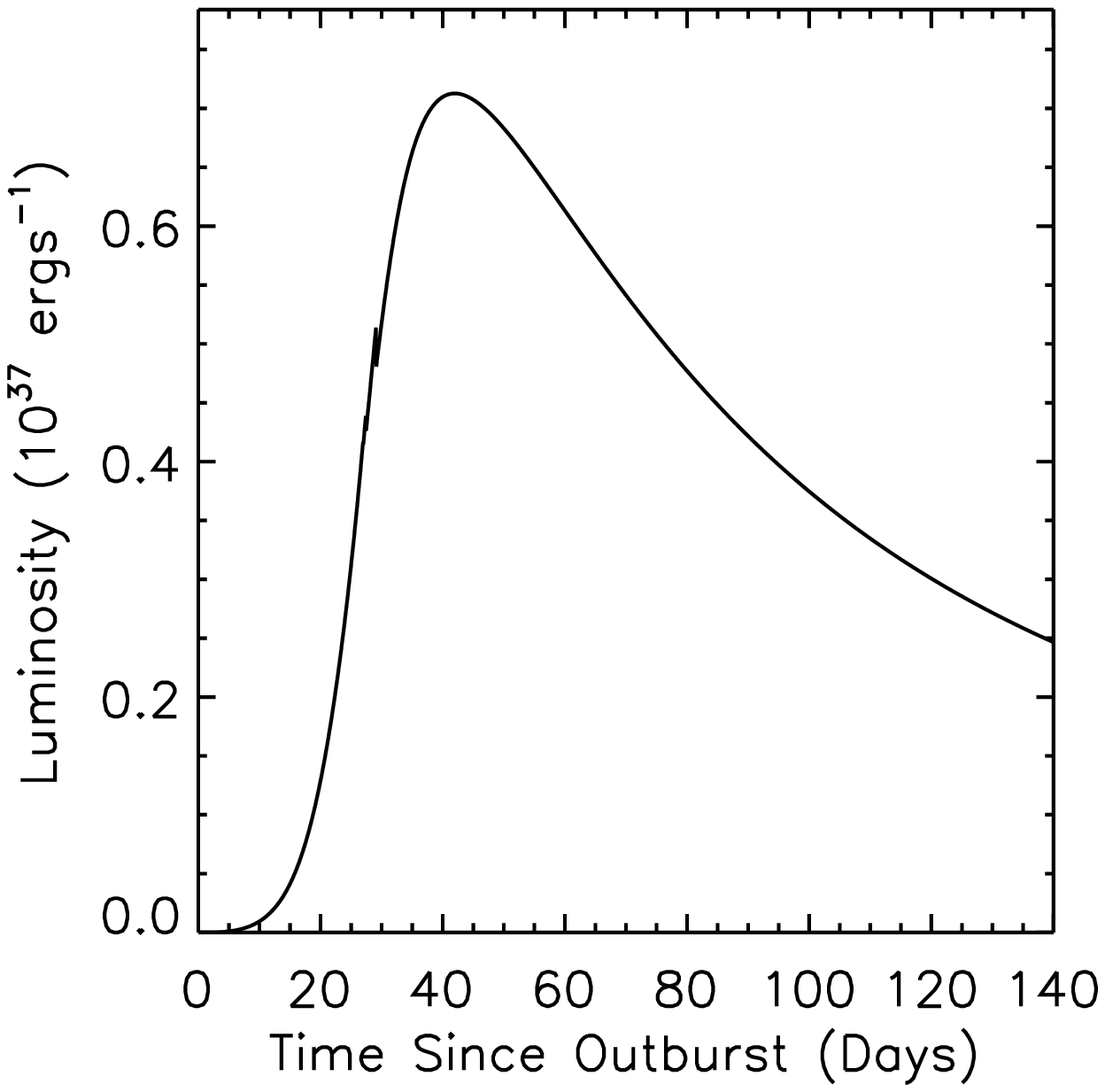,width=39mm,angle=90} \\
\hspace{1mm}
\psfig{figure=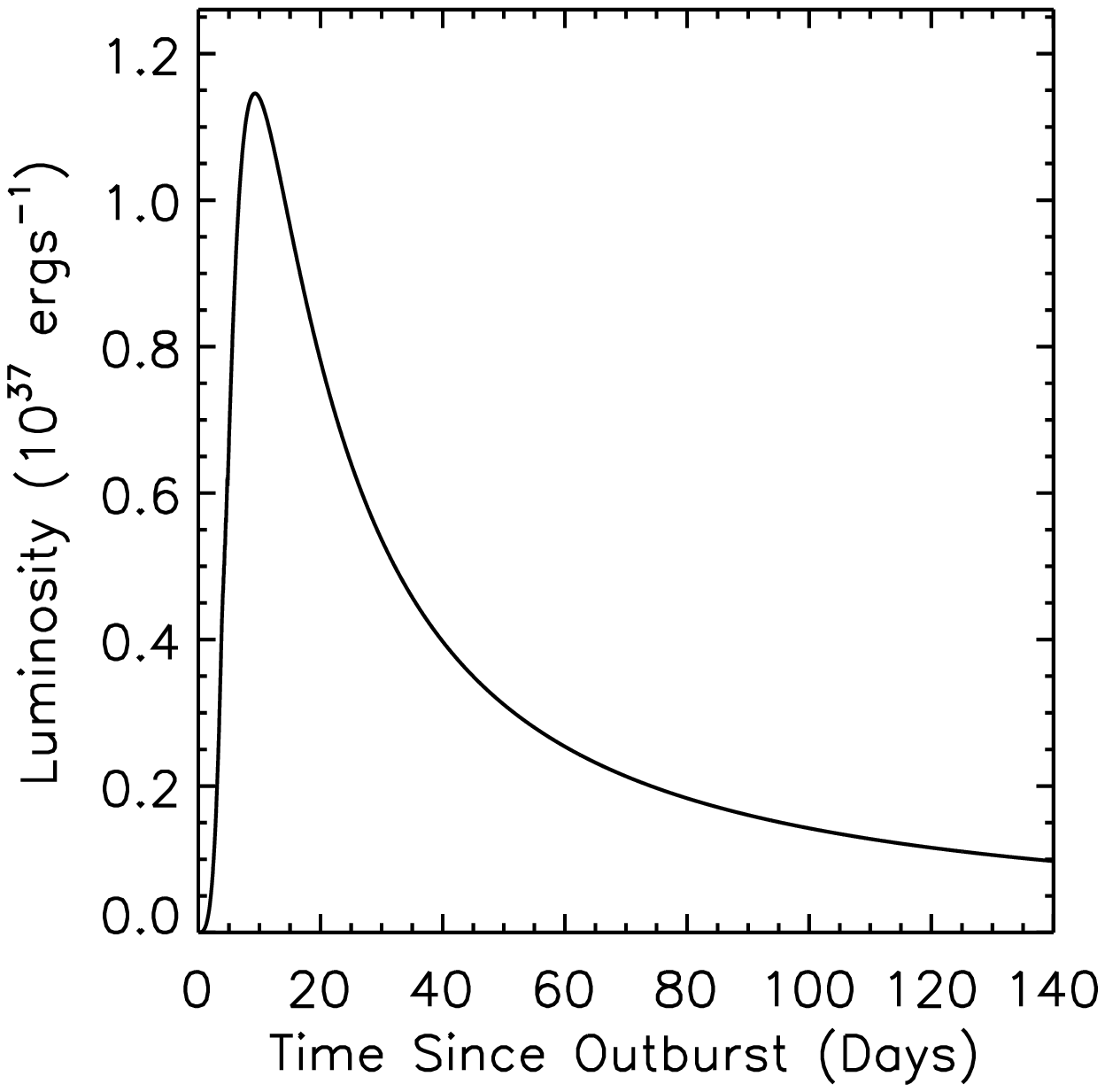,width=39mm,angle=90} &
\hspace{-2mm}
\psfig{figure=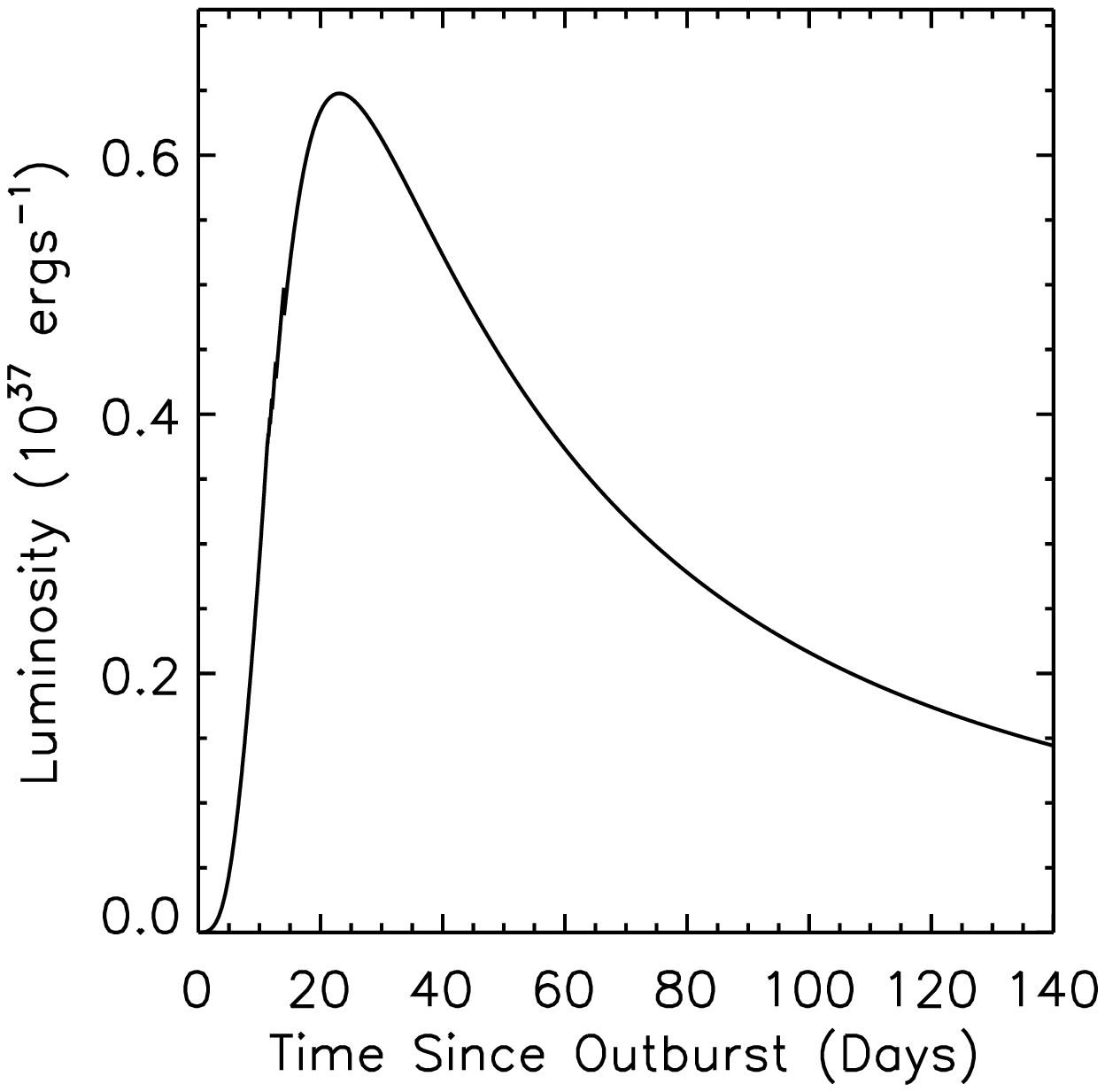,width=39mm,angle=90}
&
\hspace{-2mm}
\psfig{figure=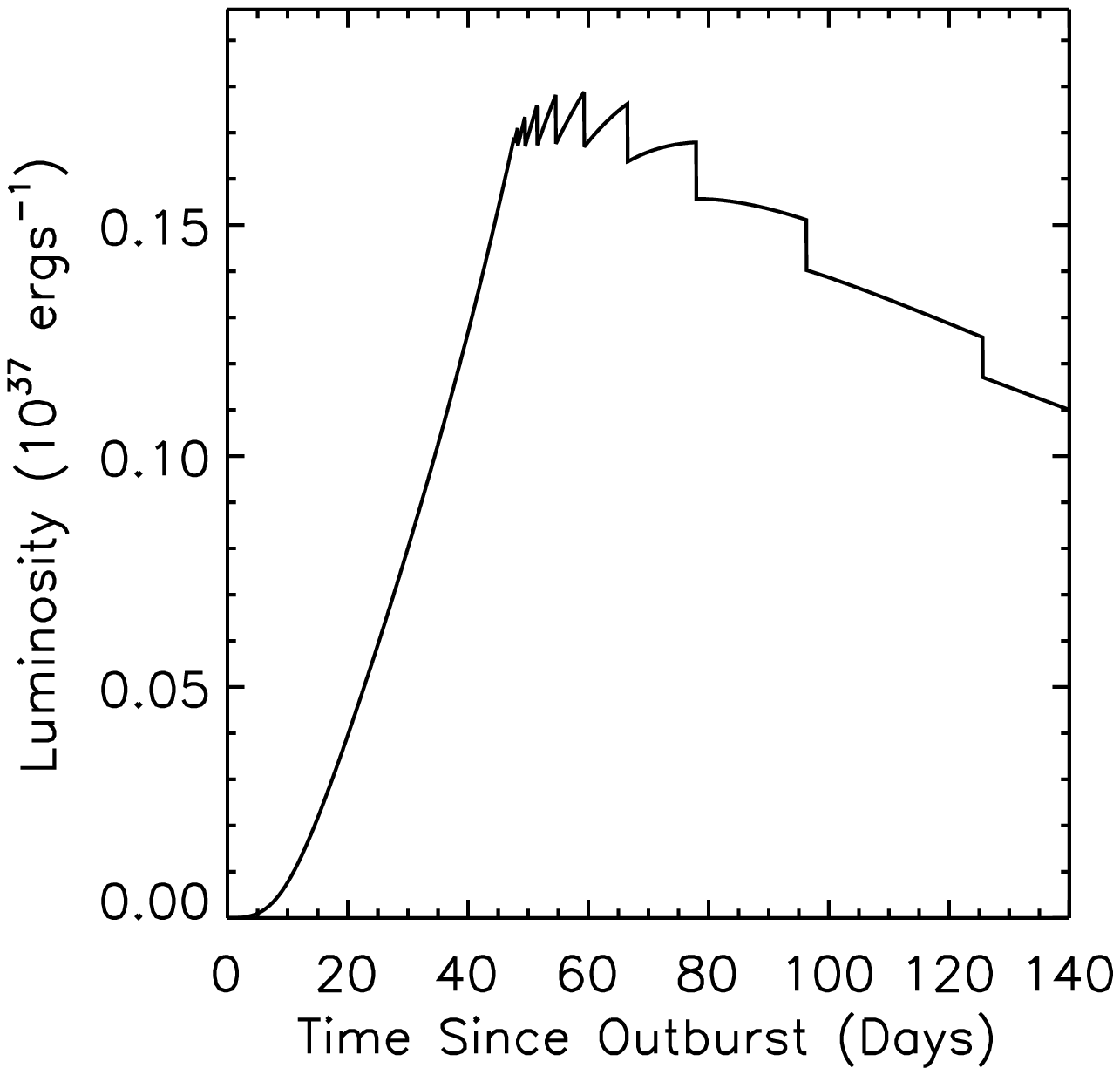,width=39mm,angle=90} \\
\hspace{1mm}
\psfig{figure=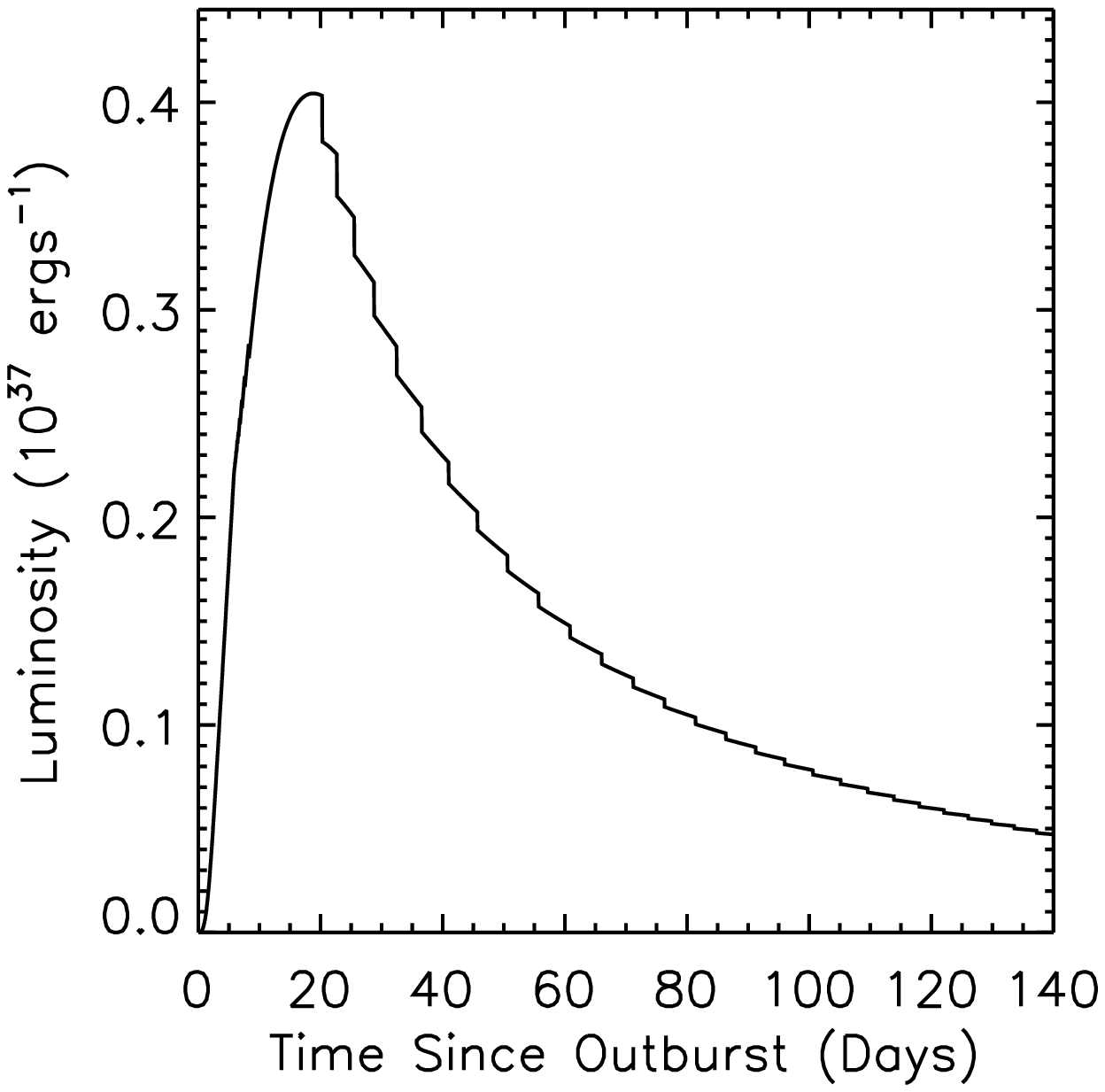,width=39mm,angle=90} &
\hspace{-2mm}
\psfig{figure=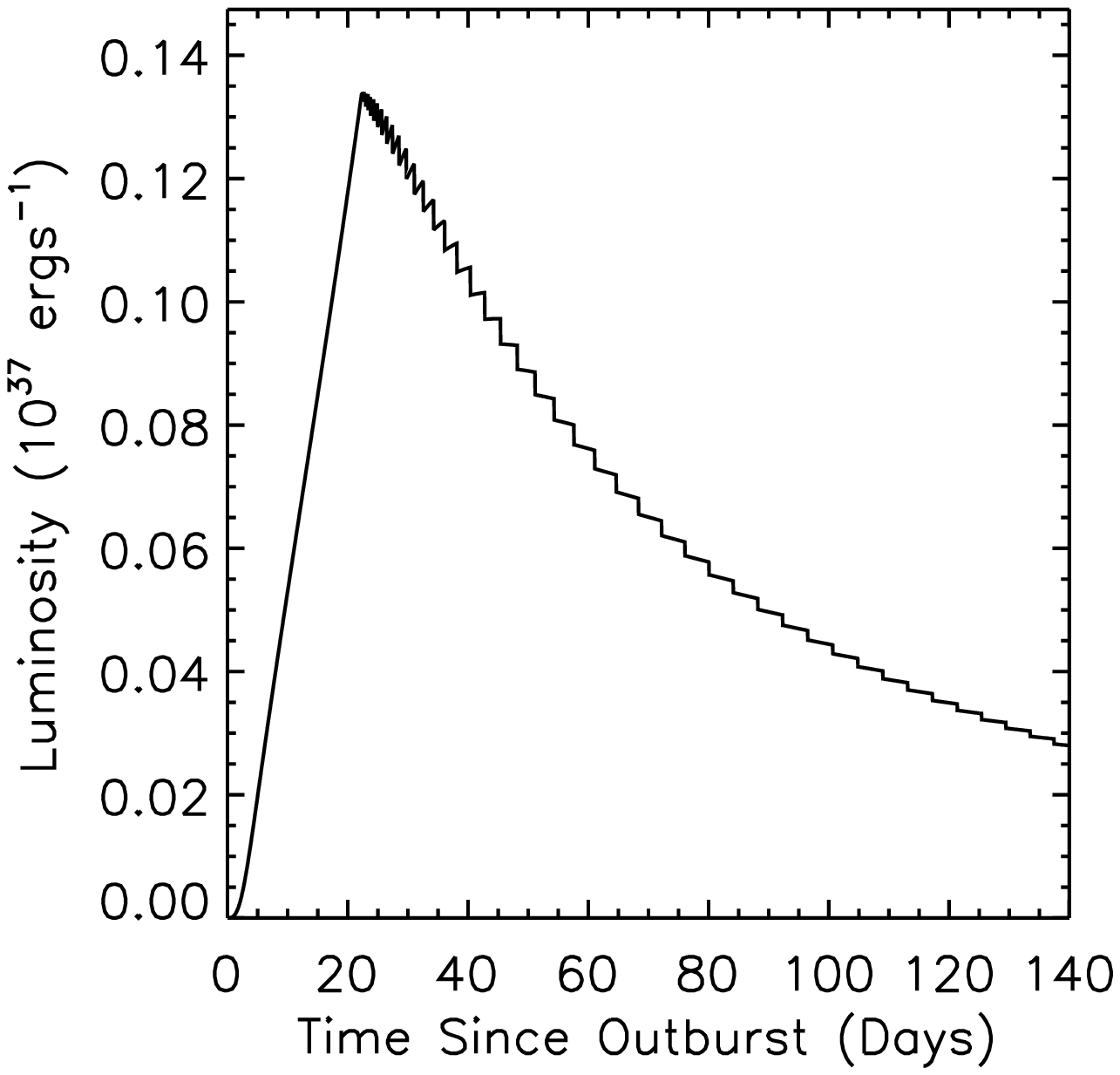,width=39mm,angle=90} &
\hspace{-2mm}
\psfig{figure=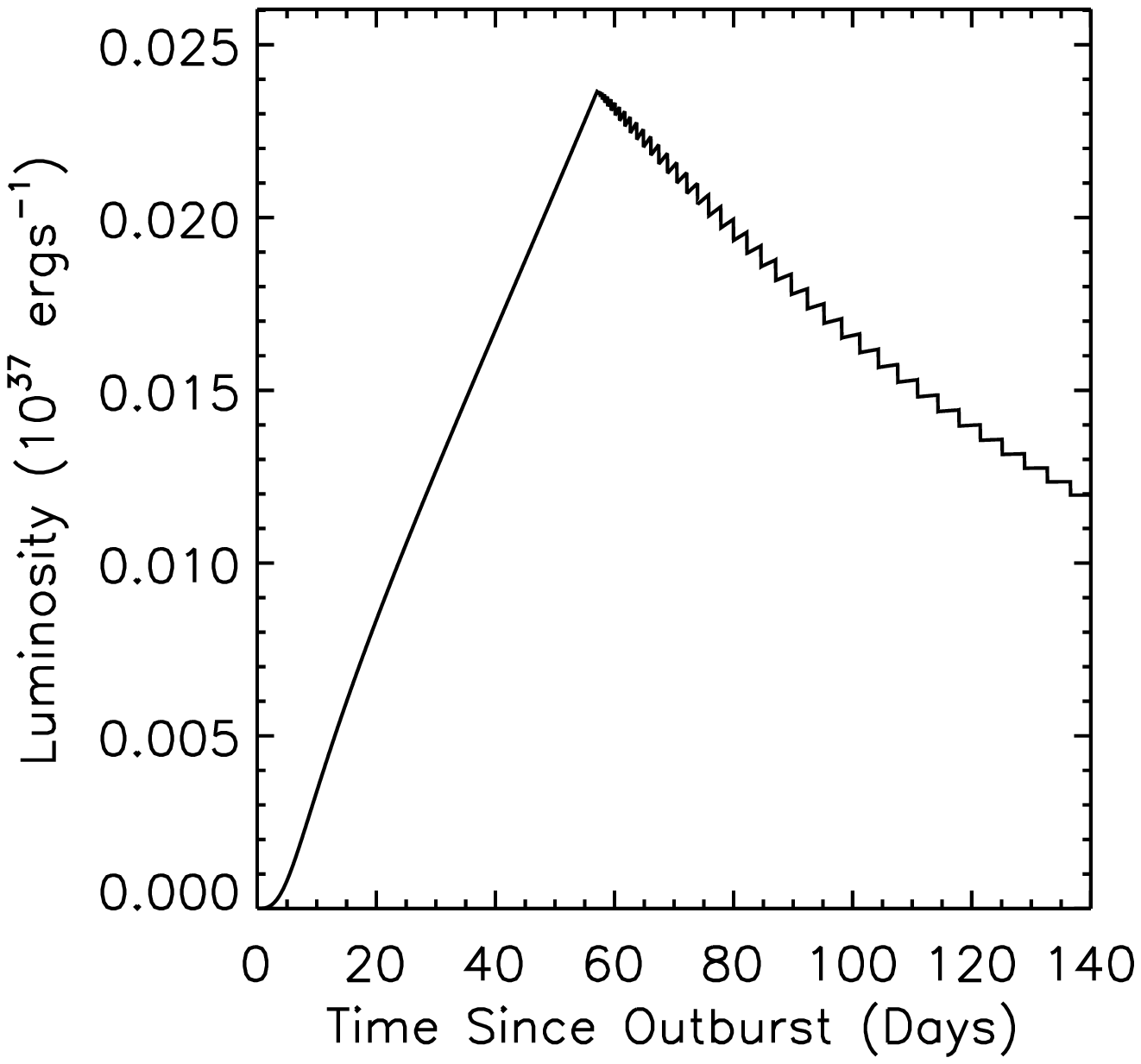,width=39mm,angle=90}
\end{tabular}}
\vspace{1mm}
\captionb{4}
{Same as Figure 3, but with a Mira mass loss rate of 10$^{-6}$
  M$_\odot$\,yr$^{-1}$.}
}
\end{figure}


Given this situation, the evolution of the shock is strongly asymmetric.  Ejecta
traveling towards the Mira encounter an increasingly high density and
evolve quickly, until they reach the photosphere.  Ejecta traveling
away from the Mira encounter progressively lower density and evolve
much more slowly.  Our model divides the ejecta into many azimuthal
bins, and tracks the interaction with the environment separately for
each bin.

The final unknown is the total ejecta mass, for which we adopted
10$^{-7}$, 10$^{-6}$, and 10$^{-5}$ M$_\odot$.   We show the
predicted luminosity evolution for these 3$\times$3$\times$3
cases in Figures 3--5.


\begin{figure}[!tH]
\vbox{
\centerline{
\begin{tabular}{ccc}
\hspace{1mm}
\psfig{figure=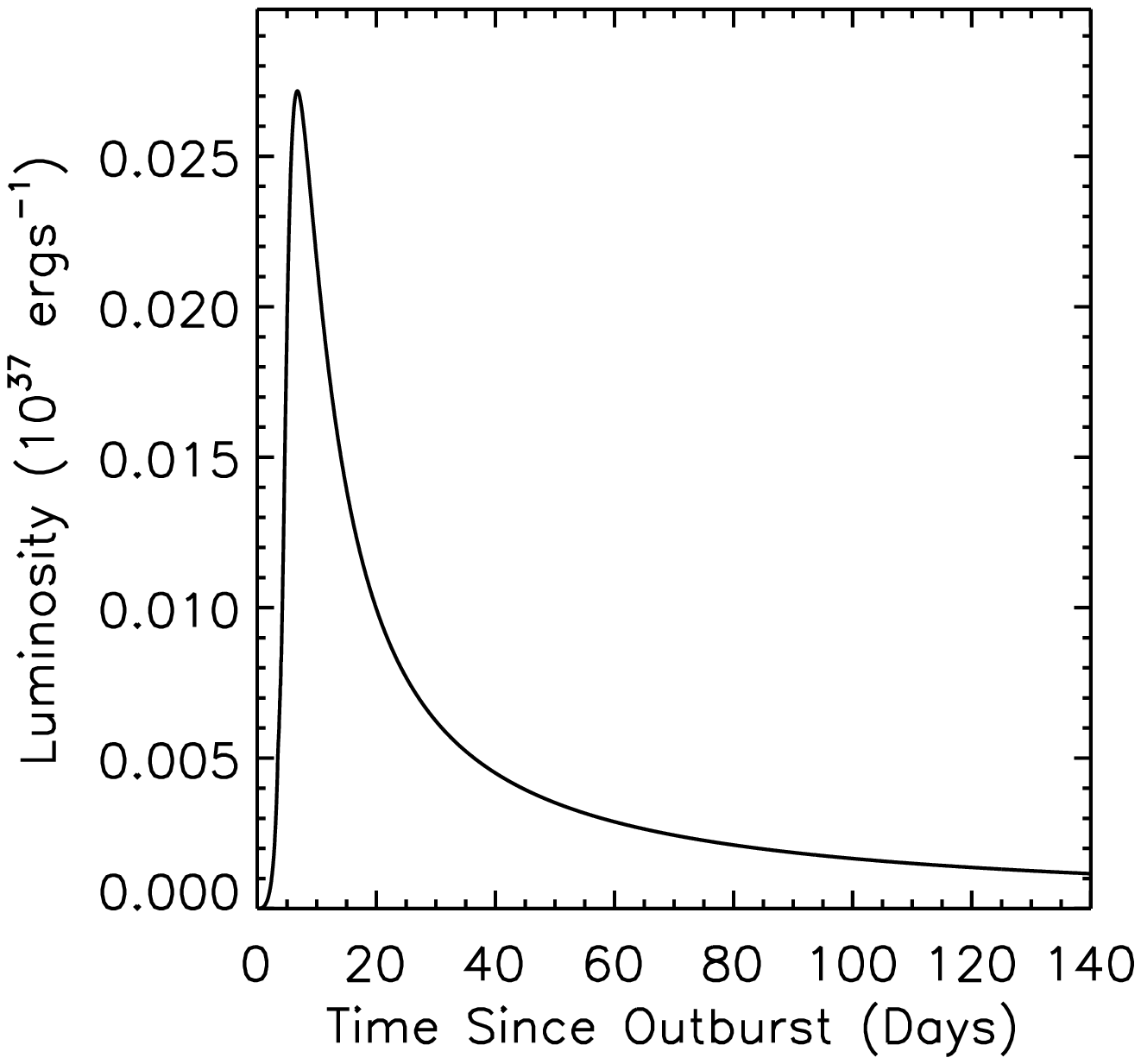,width=39mm,angle=90} &
\hspace{-2mm}
\psfig{figure=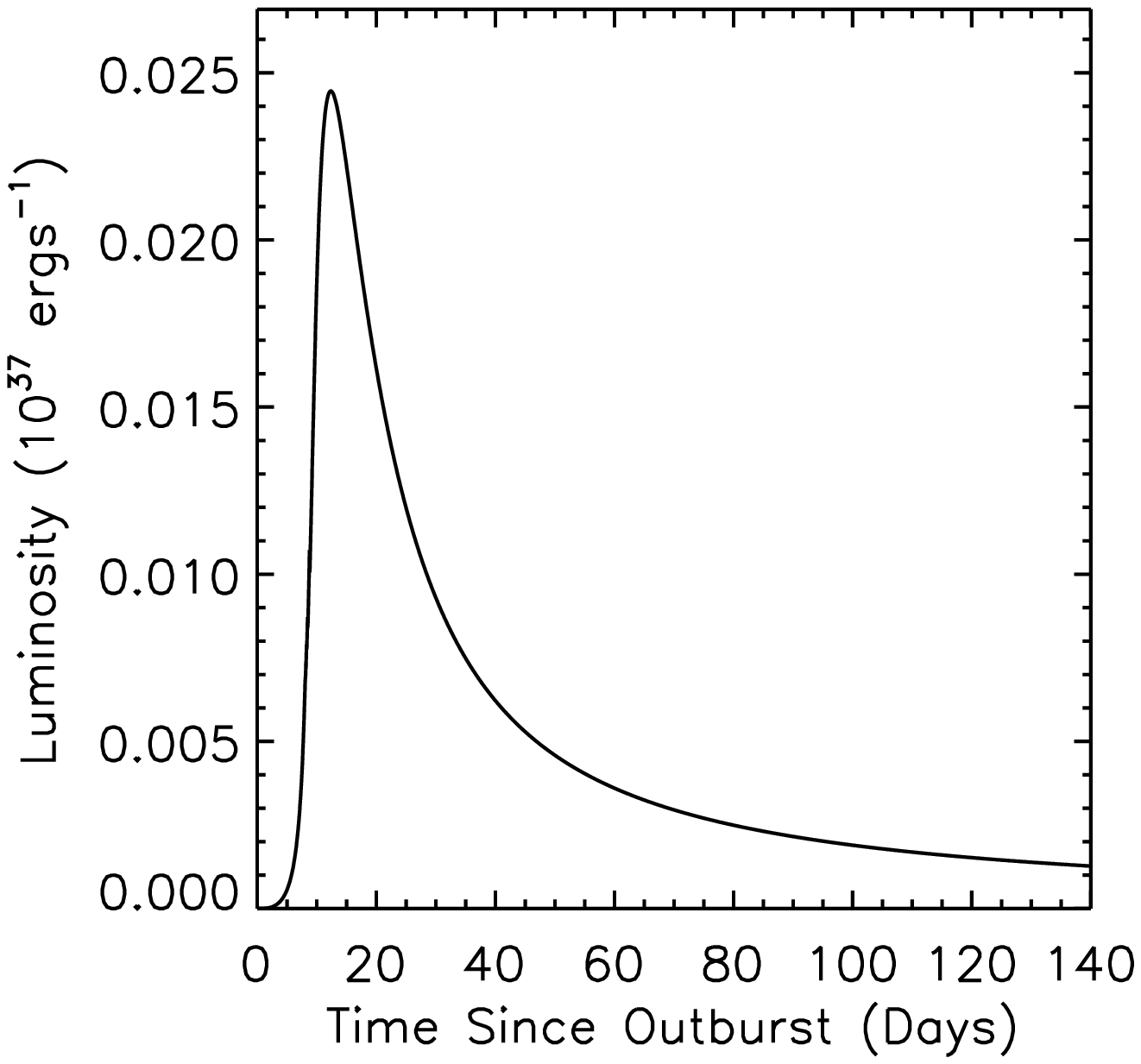,width=39mm,angle=90} &
\hspace{-2mm}
\psfig{figure=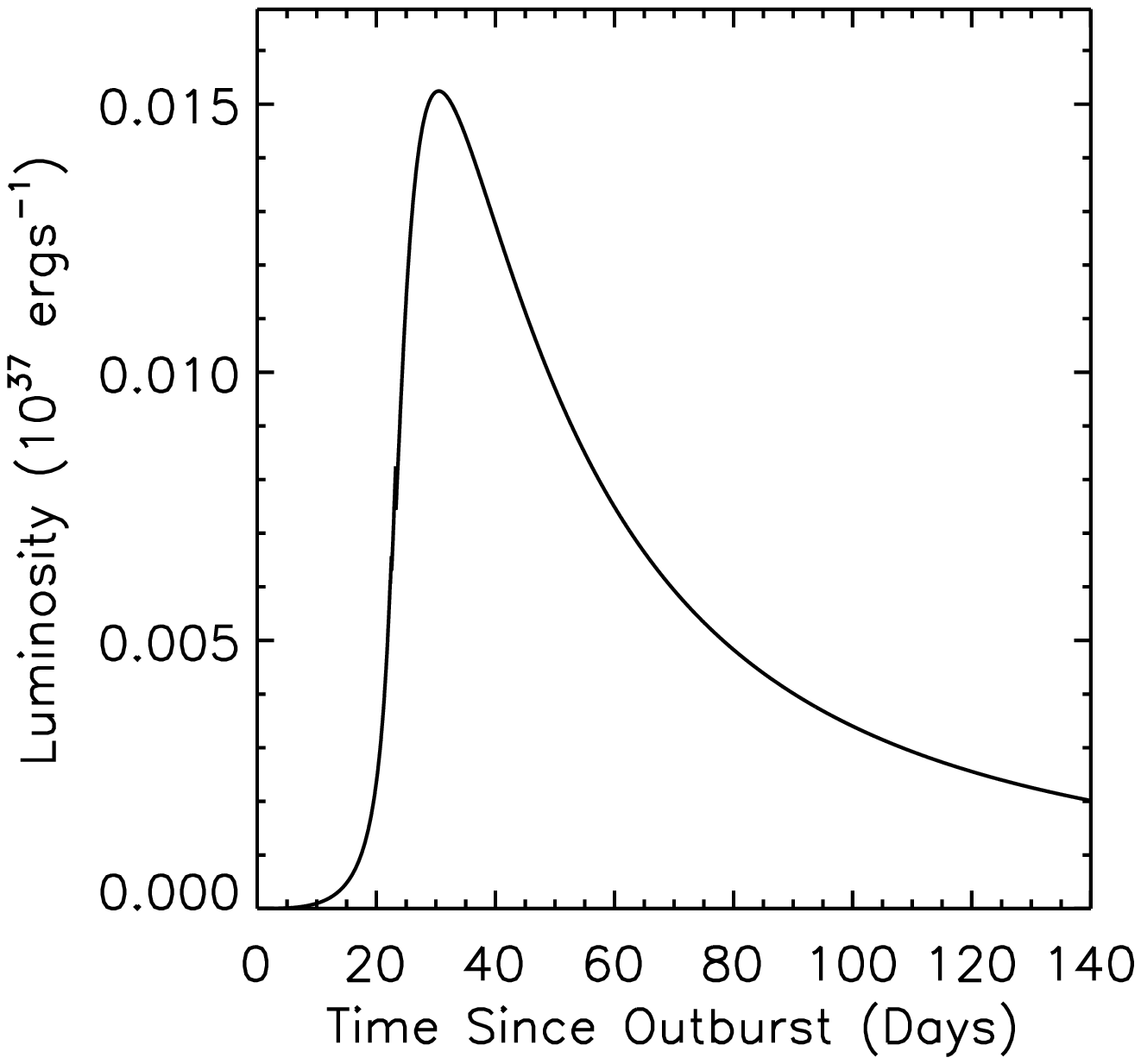,width=39mm,angle=90} \\
\hspace{1mm}
\psfig{figure=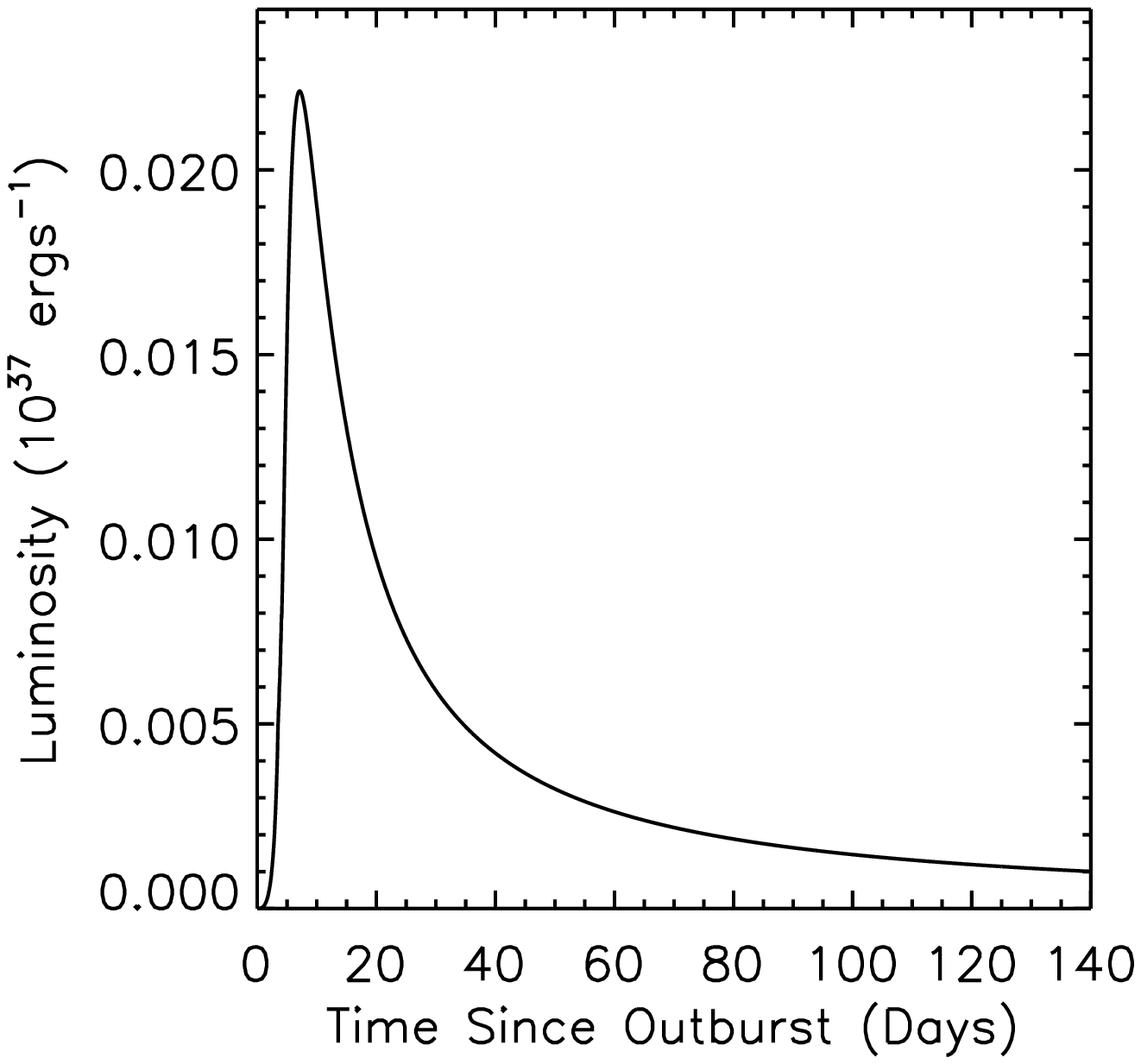,width=39mm,angle=90} &
\hspace{-2mm}
\psfig{figure=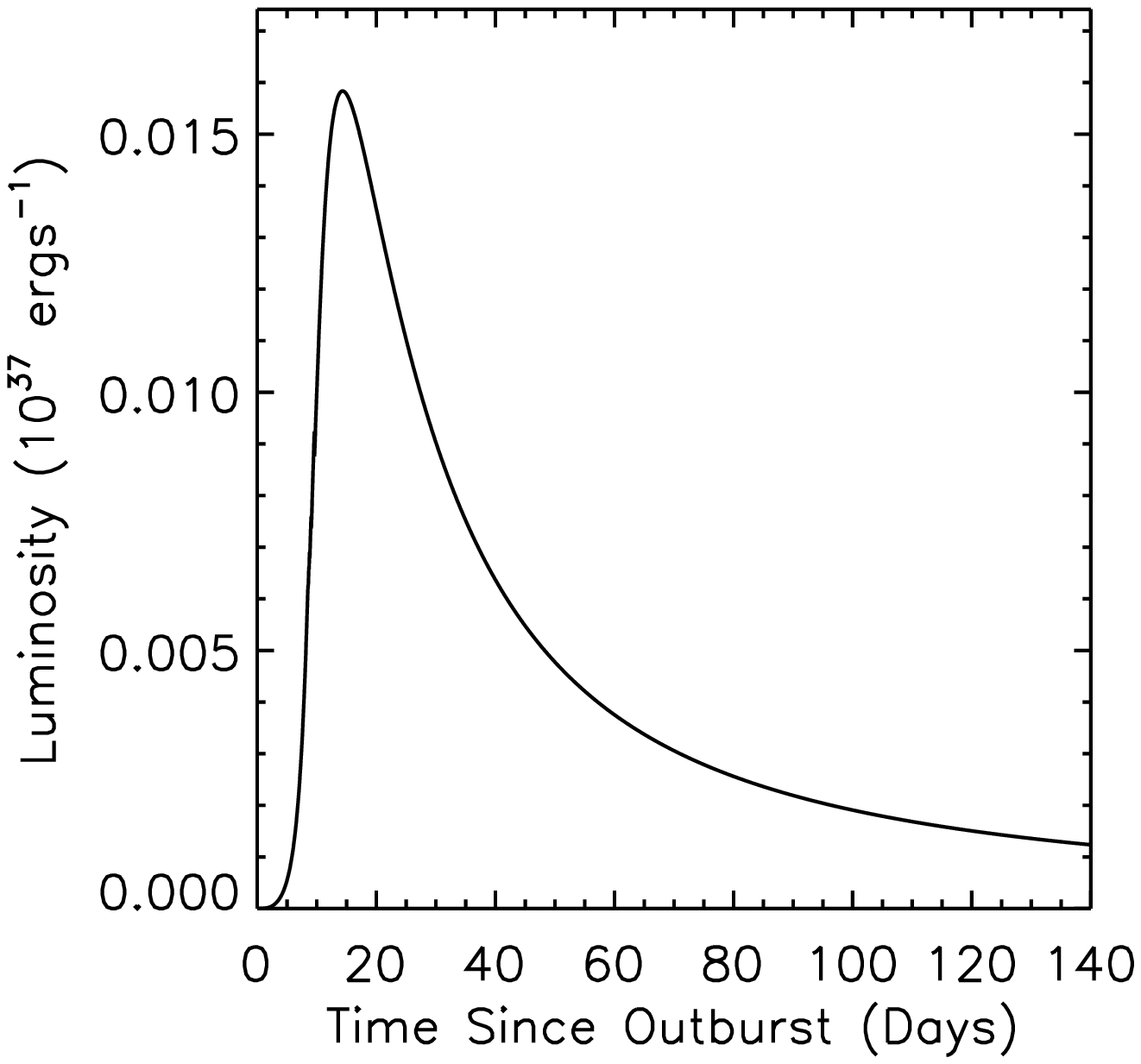,width=39mm,angle=90}
&
\hspace{-2mm}
\psfig{figure=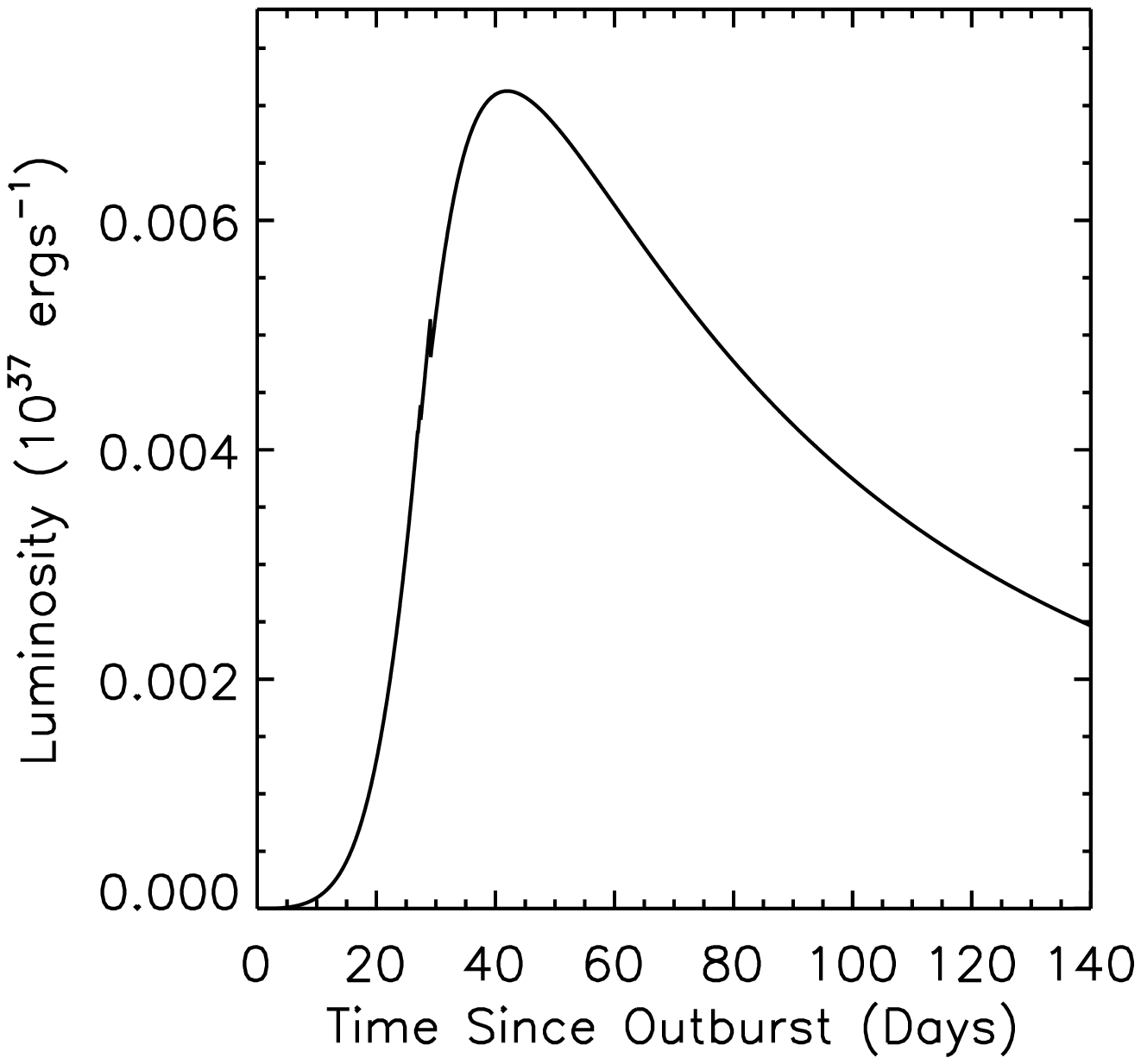,width=39mm,angle=90} \\
\hspace{1mm}
\psfig{figure=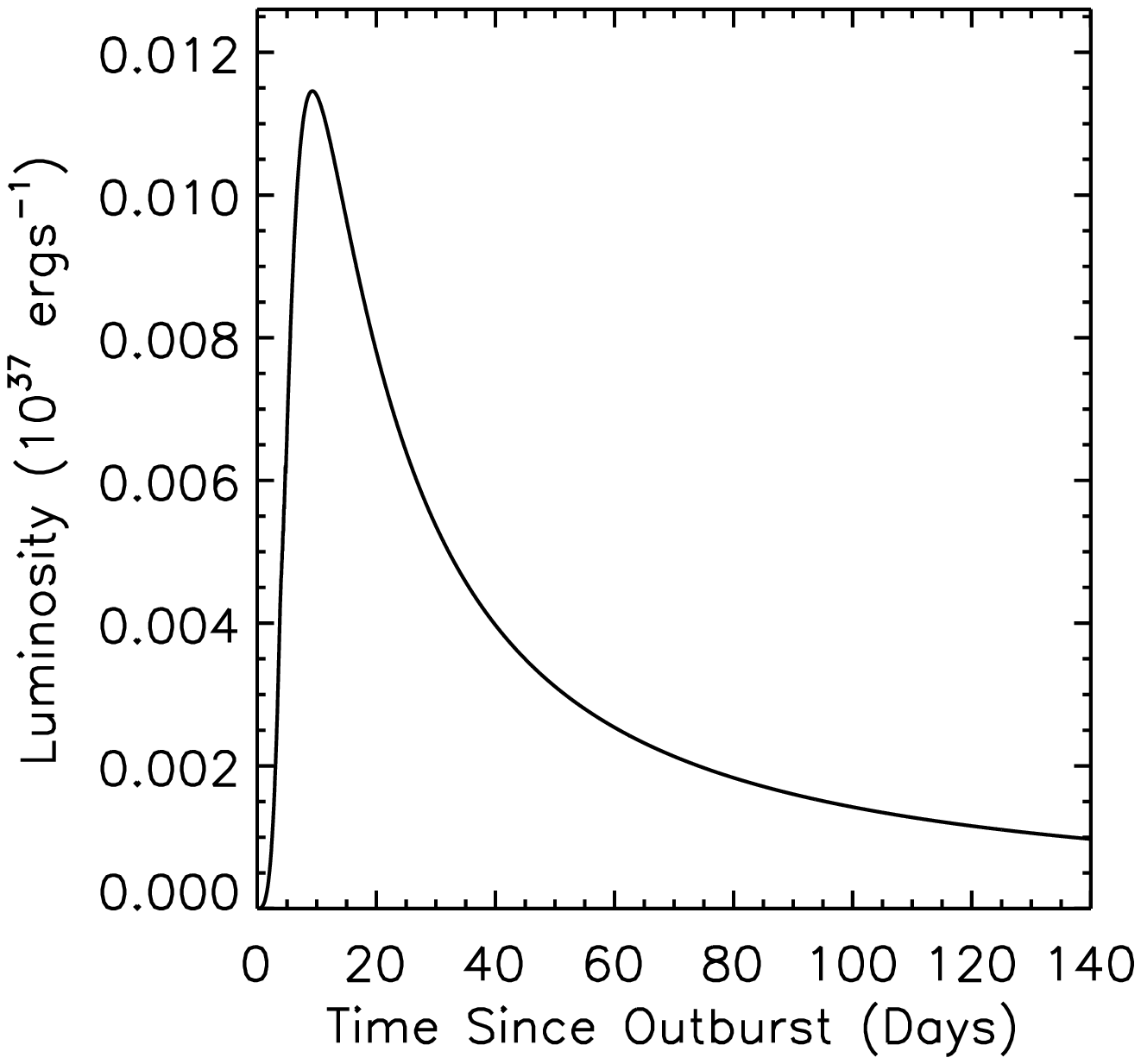,width=39mm,angle=90} &
\hspace{-2mm}
\psfig{figure=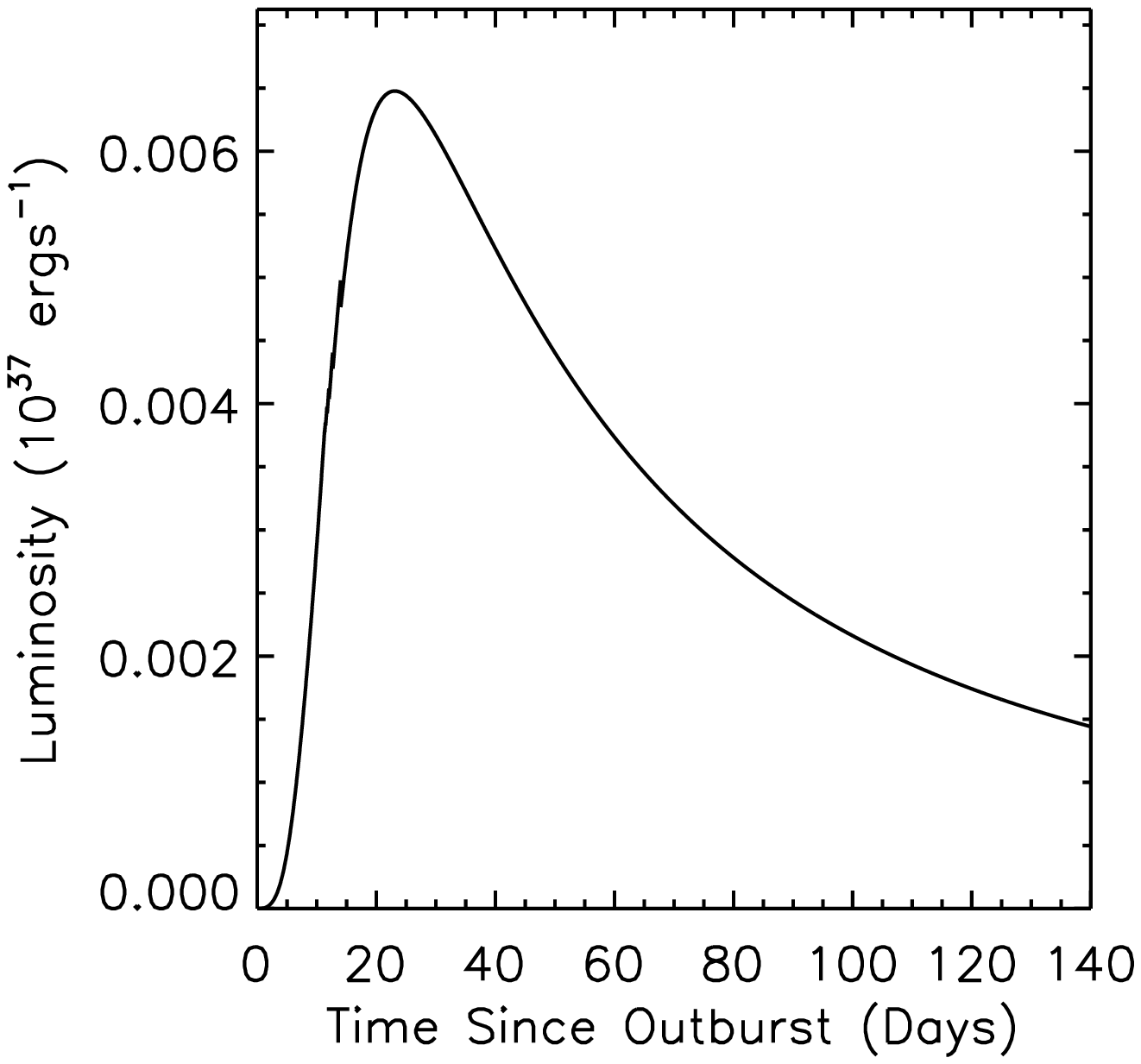,width=39mm,angle=90} &
\hspace{-2mm}
\psfig{figure=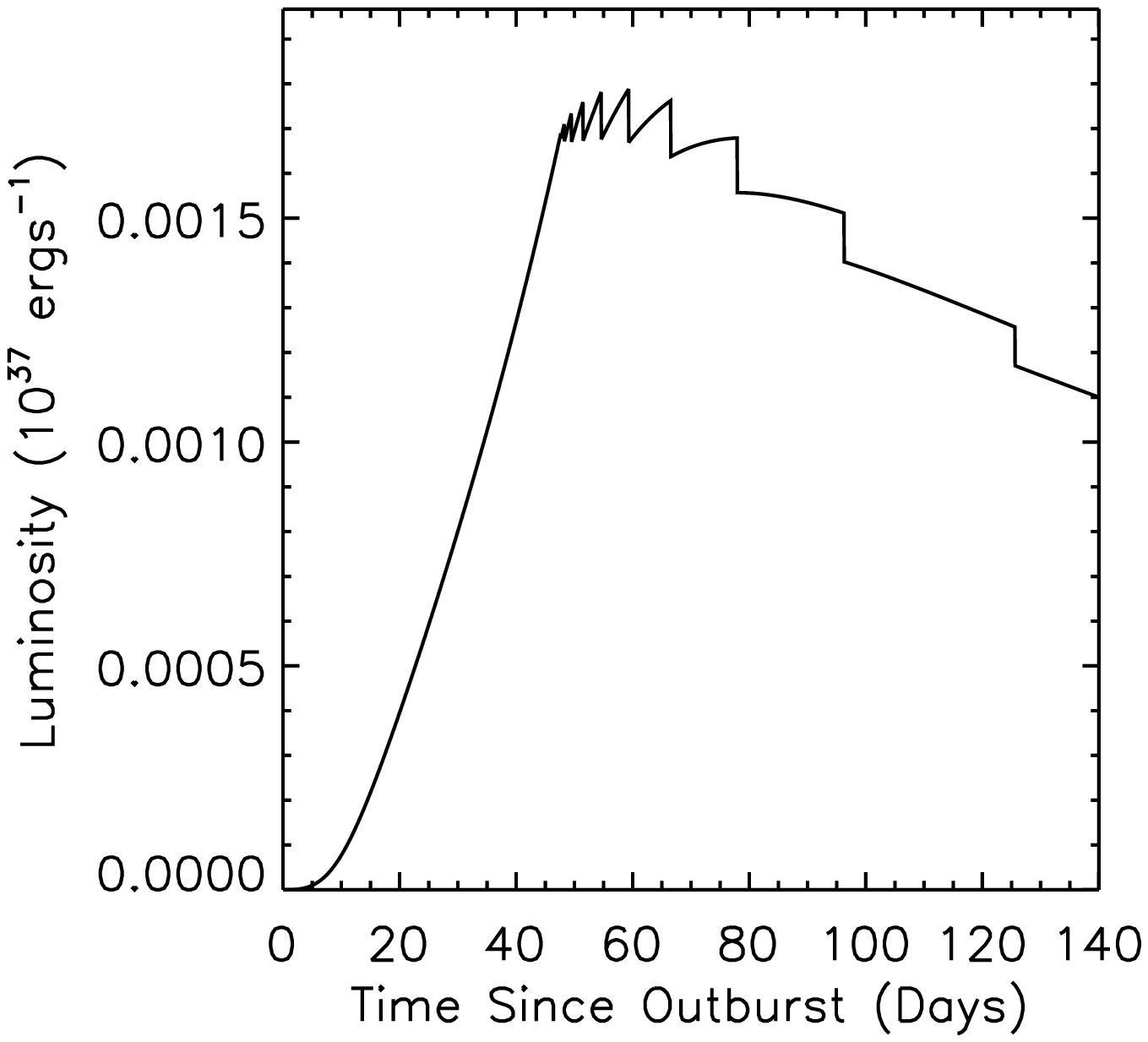,width=39mm,angle=90}
\end{tabular}}
\vspace{1mm}
\captionb{5}
{Same as Figure 3, but with a Mira mass loss rate of 10$^{-7}$
  M$_\odot$\,yr$^{-1}$.}
}
\end{figure}


We first note that this model can reproduce the shape of the observed
X-ray light curve using certain combinations of the parameters (e.g.,
a binary separation of 50 AU, ejecta mass of 10$^{-5}$ M$_\odot$, and
Mira wind mass loss rate of 10$^{-7}$ M$_\odot$\,yr$^{-1}$).  That is,
our model predicts an initial phase of low flux, a sudden increase in
luminosity, and a decline with a changing slope.

Based on this success, we identify the sudden brightening observed
around day 20 with the time when the blast wave reached the immediate
vicinity of the Mira.  To first order, this implies a binary
separation of 40 AU for an ejecta velocity of 3200 km\,s$^{-1}$.  The real separation is
smaller by an amount dependent on the degree of deceleration during
this period.  However, significant deceleration necessarily implies
significant interaction, which results in increased X-ray emission
well before the ejecta reach the Mira surface.  Thus, parameter
combinations that have low ejecta mass and/or high wind mass loss
rates result in a gradual increase in X-ray luminosity, compared to
the sudden increase that is observed.

Thus, our model results suggest a large binary separation and a modest
mass loss rate for the Mira in V407 Cyg.   However, it should be kept
in mind that ours is a simple model.  A full physical model of the
ejecta--wind interaction might result in a somewhat different
conclusion.  We also suggest that a careful modeling of the existing body
of optical spectra during the first 20 days should be performed to
infer the degree of deceleration towards of the ejecta traveling
towards the Mira.  This could provide the best measurement of
the binary separation, assuming our interpretation of the sudden X-ray
brightening is correct.

\sectionb{4}{CONCLUSIONS}

Our analysis suggests that V407 Cyg harbors a massive white dwarf in a
wide  binary, much wider than RS Oph.  The blast wave initially
encounters a relatively low density ($< 10^6$ cm$^{-3}$) stellar wind.  The subsequent
evolution is highly asymmetric; in one direction, the wind density
sharply increases around day 20 as the ejecta approach the Mira
surface.  This leads to the sudden increase in the X-ray emission.  In
the opposite direction, the wind density decreases.  This asymmetry
results in an X-ray spectrum that can be modeled using our CIE+NIE,
two-component model.

The GeV $\gamma$-ray emission detected with {\sl Fermi\/}/LAT (Abdo
et al. 2010) requires both particle acceleration and a secondary
process that generates the $\gamma$-rays.  The latter may either be
accelerated protons hitting the Mira, producing pions, or the infrared
photons of the Mira inverse-Compton scattering off relativistic
electrons.  In the direction away from the Mira, both the number of
accelerated particles and that of the target particles continue to
decrease with time.  In the direction towards the Mira, the 
deceleration caused by the high density environment leads to
a sharp decrease in the maximum energy that the particles can be
accelerated to, even though the density of seed and target particles
both increase.  This, in our view, is the cause of the drop in
$\gamma$-ray intensity at the time of X-ray flux increase.

The exploration of parameter space of our simple model suggests
that the binary separation is larger ($>$20 AU) than
the proposed orbital period of 43 years would imply.  It also suggests
a modest wind mass loss rate (not much more than 10$^{-7}$
  M$_\odot$\,yr$^{-1}$ ) for the Mira.  These tentative results
should be checked with additional observations.

V407 Cyg joins a small group of symbiotic stars that have exhibited a
fast nova outburst.  Other four are all known recurrent
novae, and V407 Cyg may yet prove to be one.  The short duration of
the supersoft phase suggests that its white dwarf may be massive
enough to have a short recurrence time.  It is quite possible,
therefore, that V407 Cyg in quiescence is a luminous, hard X-ray
source similar to T CrB and several other symbiotic stars (Kennea et
al. 2009).  On the other hand, RS Oph in quiescence is significantly fainter
than T CrB (Nelson et al. 2011).   We believe that a quiescent X-ray
observation should be attempted as a matter of priority.

\References

\refb Abdo, A.A., Ackermann, M., Ajello, M. et al. 2010, Science, 329, 817

\refb Bode, M.F., O'Brien, T.J., Osborne, J.P. et al. 2006, ApJ, 652,
629

\refb Kennea, J.A., Mukai, K., Sokoloski, J.L. et al. 2009, ApJ, 701,
1992

\refb Munari, U., Margoni, R., Stagni, R. 1990, MNRAS, 242, 653

\refb Munari, U., Joshi, V.H., Ashok, N.M. et al. 2011, MNRAS, 410,
L52

\refb Nelson, T., Orio, M., Cassinelli, J.P. et al. 2008, ApJ, 673,
1067

\refb Nelson, T., Mukai, K., Orio, M. et al. 2011, ApJ, 737, 7

\refb Nelson, T., Donato, D., Mukai, K. et al. 2012, ApJ, in press

\refb Ness, J.-U.,Drake, J.J., Starrfield, S., et al. 2009, AJ, 137, 3414

\refb Orlando, S., Drake, J.J., Laming, J.M. 2009, A\&A, 493, 1049

\refb Osborne, J.P., Page, K.L., Beardmore, A.P. et al. 2011, ApJ,
727, 124

\refb Shore, S.N., Wahlgren, G.M., Augusteijn, T. et al. 2011,
A\&A, 527, A98

\refb Smith, R.K., Hughes, J.P. 2010, ApJ, 718, 583

\refb Sokoloski, J.L., Luna, G.J.M., Mukai, K., Kenyon, S.J. 2006,
Nature, 442, 276

\refb Zajczyk, A., Tomov, T., Miko\l{}ajewski, M. et al., 2007, Baltic
Astronomy, 16, 62

\end{document}